%% file: main.tex
\documentclass[manuscript, acmlarge, screen]{acmart}

\AtBeginDocument{
  }

\setcopyright{cc}
\setcctype{by}
\acmJournal{TOSEM}
\acmYear{2026}
\acmVolume{1}
\acmNumber{1}
\acmArticle{}
\acmMonth{1}
\acmDOI{10.1145/3842759}

\input{setup}

\begin{document}

\begin{CCSXML}
  <ccs2012>
  <concept>
  <concept_id>10011007.10011006.10011073</concept_id>
  <concept_desc>Software and its engineering~Software maintenance tools</concept_desc>
  <concept_significance>500</concept_significance>
  </concept>
  </ccs2012>
\end{CCSXML}

\ccsdesc[500]{Software and its engineering~Software maintenance tools}

\keywords{Log analysis, Software logs, Large language models}

\title{LLM4Log: A Systematic Review of Large Language Model-based Log Analysis}

\author{Zeyang Ma}
\orcid{0000-0002-0390-1547}
\affiliation{
  \institution{Software PErformance, Analysis, and Reliability (SPEAR) lab, Concordia University}
  \city{Montreal}
  \country{Canada}
}
\email{m_zeyang@encs.concordia.ca}
\author{Jinqiu Yang}
\orcid{0000-0003-4282-406X}
\affiliation{
  \institution{Department of Computer Science and Software Engineering, Concordia University}
  \city{Montreal}
  \country{Canada}
}
\email{jinqiu.yang@concordia.ca}
\author{Tse-Hsun (Peter) Chen}
\orcid{0000-0003-4027-0905}
\affiliation{
  \institution{Software PErformance, Analysis, and Reliability (SPEAR) lab, Concordia University}
  \city{Montreal}
  \country{Canada}
}
\email{peterc@encs.concordia.ca}

\pagestyle{plain}

\input abstract
\maketitle

\input introduction

\input Survey_methodology

\input background

\input Logging

\input Log_Parsing

\input Log_Mining

\input Cross_Cutting_Insights

\input conclusion

\balance
\bibliographystyle{plainnat}
\bibliography{ref}
\end{document}

%% file: setup.tex
\usepackage{listings}
\usepackage{tcolorbox}
\usepackage{soul}
\usepackage{xcolor}
\usepackage{fancybox}
\usepackage{xspace}
\usepackage{ragged2e}
\usepackage{changepage}
\usepackage{makecell}
\usepackage{setspace}
\usepackage{enumitem}
\usepackage{booktabs, multirow} 

\usepackage{url}

\usepackage{hyperref}
\hypersetup{colorlinks=false}

\definecolor{codebg}{RGB}{248,248,248}
\definecolor{codekw}{RGB}{0,92,197}
\definecolor{codestr}{RGB}{163,21,21}
\definecolor{codecmt}{RGB}{0,128,0}
\definecolor{codeln}{RGB}{120,120,120}

\lstdefinestyle{logcode}{
  backgroundcolor=\color{codebg},
  basicstyle=\ttfamily\tiny,
  keywordstyle=\color{codekw}\bfseries,
  stringstyle=\color{codestr},
  commentstyle=\color{codecmt}\itshape,
  numbers=left,
  numberstyle=\tiny\color{codeln},
  numbersep=6pt,
  stepnumber=1,
  showstringspaces=false,
  breaklines=true,
  frame=single,
  rulecolor=\color{black!20},
  tabsize=2,
  captionpos=b
}

\usepackage{amsmath,amssymb,amsfonts}

\usepackage{textcomp}
\usepackage{lipsum}
\usepackage{graphicx}
\usepackage{tikz}
\usetikzlibrary{positioning,arrows.meta,shapes.misc,fit,calc}
\usepackage{subfigure}
\usepackage{balance}
\usepackage{wrapfig}
\usepackage{needspace}
\usepackage[ruled, lined, linesnumbered, commentsnumbered, longend]{algorithm2e}
\usepackage{algorithmic}
\SetCommentSty{mycommfont}
\usepackage{array}

\usepackage{adjustbox}
\usepackage{threeparttable}

\newcommand{\phead}[1]{\vspace{1mm} \noindent {\bf #1}}
\newcolumntype{?}{!{\vrule width 1.5pt}}

\newcommand{\rqboxc}[1]{\begin{tcolorbox}[left=1pt,right=1pt,top=1pt,bottom=1pt,colback=gray!5,colframe=gray!40!black,before skip=5pt,after skip=0pt]#1\end{tcolorbox}}

\definecolor{lightgray}{gray}{0.9}

%% file: abstract.tex
\begin{abstract}
  Software systems generate massive, evolving, semi-structured logs that are central to reliability engineering and AIOps, yet difficult to analyze at scale under drift and limited labels. Recent advances in pretrained Transformer models and instruction-tuned large language models (LLMs) have reshaped log analysis by enabling semantic generalization and cross-source evidence integration, but also introducing deployment risks such as context limits, latency and cost, privacy constraints, and hallucinations.
  This paper presents \textbf{LLM4Log}, a systematic review of LLM-based log analysis across the end-to-end pipeline, from upstream logging-statement generation and maintenance to log parsing/structuring and downstream tasks including anomaly detection, failure prediction, root cause analysis, and log summarization. Following a structured search and manual screening protocol, we completed literature collection in November 2025 and identified 145 unique papers across seven logging tasks. We organize the research area through a unified, task-driven taxonomy, summarize common design patterns (prompting/ICL, retrieval grounding, fine-tuning, tool/agent augmentation, and verification), and analyze evaluation practices, datasets, metrics, and reproducibility.
  Based on these cross-paper analyses, we summarize key lessons and open challenges for reliable real-world adoption. We emphasize robustness under drift and long-tail events, grounding and faithfulness for operator-facing outputs, and deployment-oriented designs with verifiable behavior.
\end{abstract}

%% file: introduction.tex
\section{Introduction}
\label{sec:introduction}

Software systems, from cloud services and microservices to networked and cyber-physical infrastructures, emit massive volumes of execution logs that record runtime states, events, and failures.
Logs are a primary source of evidence for reliability engineering and operations: they support debugging, monitoring, incident triage, and postmortem analysis~\cite{fu2014wheredolog,zhu2015learningtolog,Yang2023EmbeddedLogs}.
However, modern log streams are notoriously difficult to analyze at scale.
They are \emph{semi-structured} (mixing free-form text with identifiers and code-like tokens), \emph{high-volume/high-velocity}, and \emph{continuously evolving} due to software updates, configuration changes, and deployment diversity.
Consequently, manual inspection and keyword search remain common yet brittle, while automated pipelines often suffer from format drift, long-tail behaviors, limited labeled data, and inconsistent logging quality~\cite{he2021survey,drain2017icws,dai2020logram,Li2023LogMessageReadability,Rong2023LoggingPractices}.

To address these challenges, the research community has developed a rich body of work on automated log analysis, typically organized as a pipeline spanning logging practice, parsing/structuring, representation learning, anomaly/failure detection, failure prediction, diagnosis, and summarization~\cite{he2021survey}.
In production settings, however, log analysis is constrained by operational realities. Incidents unfold under time pressure. Log streams are interleaved across services, hosts, and tenants. Failures manifest through long causal chains. The most diagnostic events are often rare, noisy, or buried in high-volume background messages~\cite{Yang2023EmbeddedLogs,Ma2025PractitionerExpectationsLogAD}. Moreover, engineers rarely rely on logs alone. Effective triage requires correlating logs with metrics/traces, historical incidents, runbooks, and code/configuration context~\cite{Yang2023EmbeddedLogs,Ma2025PractitionerExpectationsLogAD}.
These characteristics make log analysis not only a \emph{pattern recognition} problem, but also an \emph{evidence-integration} and \emph{human decision-support} problem.
Log-centric LLM systems are therefore distinct from broader LLM4AIOps settings: logs combine natural-language-like descriptions with code-like identifiers, exhibit ambiguous template/value boundaries, evolve under version and deployment drift, and often carry the chronological symptom evidence that operators use to diagnose failures.
The challenge is not merely to classify patterns, but to interpret noisy, long-tail runtime evidence in a form that remains useful for human decision making.

The classic log-analysis pipeline was largely shaped before instruction-tuned LLMs became widely available, when methods were either rule-based or relied on fixed representations and task-specific supervision.
In recent years, there has been a rapid methodological transition: pretrained Transformer language models, including modern LLMs, are increasingly used as the backbone for log analysis.
Unlike traditional approaches that heavily depend on templates, handcrafted features, or patterns inferred from previously observed log outputs, LLMs offer semantic generalization and can condition on heterogeneous evidence (e.g., logs together with tickets, traces/metrics, runbooks, configurations, and code/configuration context), which can be especially useful in modern operational workflows~\cite{owl2024iclr,Cui2024LogEval,Karlsen2024BenchmarkingLLMsJNSM}.
In log-centric settings, this matters because the same failure may surface through heterogeneous wording across components and versions, while the most useful diagnostic evidence is often sparse, context dependent, and embedded in long chains of runtime messages instead of a single fixed template.

\begin{figure}[h!]
  \centering
  \includegraphics[width=1\columnwidth]{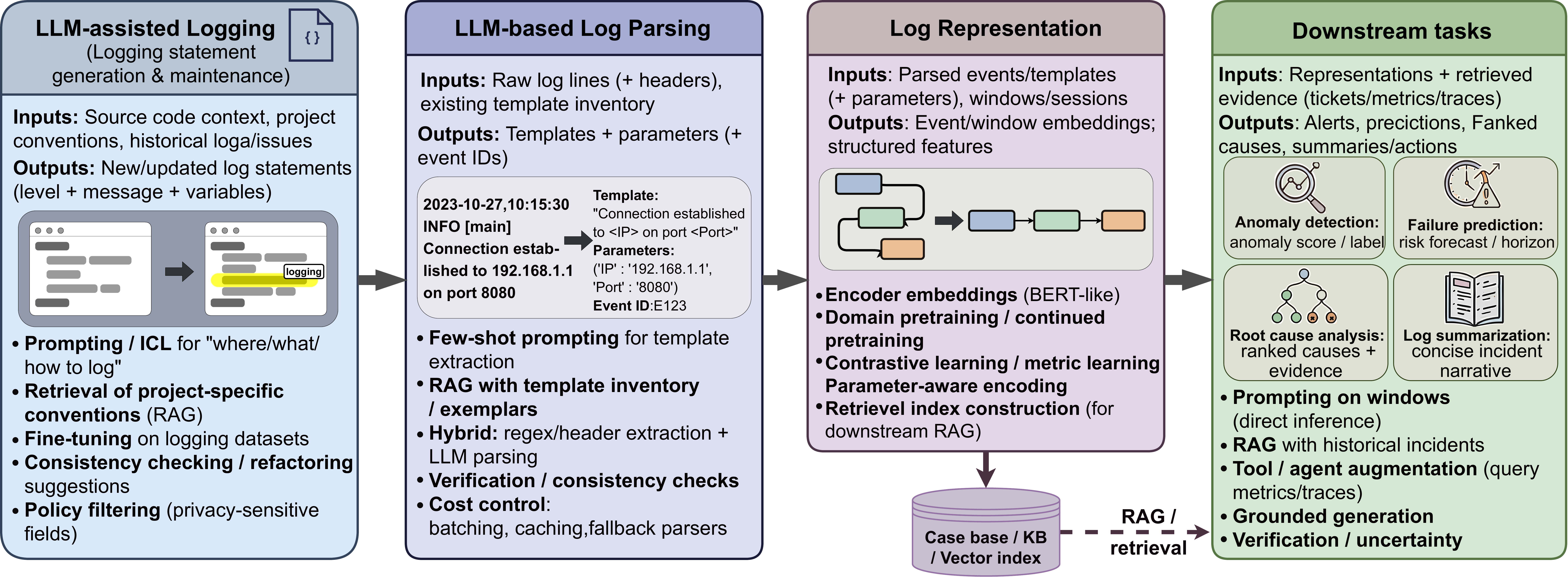}
  \caption{An overview of LLM-based log analysis across the pipeline, from logging to downstream tasks. Note that some downstream studies operate directly on raw or lightly normalized log windows and therefore bypass explicit log parsing.}
  \label{fig:overall}
\end{figure}

This shift is already visible across the \emph{entire} log-analysis pipeline.
Figure~\ref{fig:overall} provides an overview of how LLMs are applied across the log-analysis pipeline.
Upstream, LLMs are used to \textbf{generate and maintain logging statements}, helping developers decide where to log and what to record under project conventions.
At the core of many pipelines, LLMs are applied to \textbf{log parsing}, converting raw messages into structured templates and parameters with improved robustness under drift, unseen formats, and long-tail events~\cite{beck2025system}.
Between parsing and downstream inference, LLMs are also increasingly used for \textbf{log representation learning}, producing semantic embeddings or contextualized event representations for logs that improve robustness and transferability under drift.
Downstream, LLMs are used for \textbf{anomaly detection} and \textbf{failure prediction}, but also increasingly for operator-facing tasks such as \textbf{root cause analysis} (RCA) and \textbf{log summarization}, where explanation quality and evidence integration are as important as prediction accuracy~\cite{Cui2024LogEval,owl2024iclr}.
Although parsing is a common intermediate stage, some downstream studies instead operate on raw or lightly normalized logs, so the pipeline should be read as a common analytical structure, not a mandatory sequence of steps.
At the same time, LLM-based log analysis introduces new trade-offs and risks that are less prominent in classic pipelines: context window limits, latency/cost constraints in online settings, privacy and data-governance concerns, sensitivity to prompting and retrieval design, and hallucinations that can mislead incident response if outputs are not properly grounded and verified~\cite{Karlsen2024BenchmarkingLLMsJNSM,Cui2024LogEval}.

Given the pace and breadth of this emerging area, practitioners and researchers need a consolidated, task-driven view of (i) \emph{where} LLMs are applied in the log-analysis pipeline, (ii) \emph{how} they are integrated (e.g., prompting/ICL, retrieval grounding, fine-tuning, tool/agent augmentation, and verification), (iii) \emph{how} they are evaluated (datasets, metrics, baselines, and reproducibility), and (iv) \emph{what} remains open for real-world deployment.
While prior surveys have reviewed automated log analysis broadly~\cite{he2021survey} and recent work has summarized specific subareas such as LLM-based log parsing~\cite{beck2025system}, a systematic, end-to-end review of the full LLM4Log landscape is timely.

In this paper, we present \textbf{LLM4Log}, a systematic review of \emph{large-language-model-based log analysis}.
We focus on software runtime logs as the central evidence source. For downstream tasks, studies are within the log-specific scope when runtime logs are the main evaluated input or are explicitly evaluated alongside another operational data source. When log use is optional, only results that explicitly involve runtime logs support log-specific findings.
Using a task-driven search and screening protocol, we reviewed 162 task-paper records across seven log-analysis tasks, with literature collection completed in November 2025.
Deduplicating repeated task assignments by paper identity yielded 145 unique papers published from 2020 to 2025.
Building on this corpus, we organize the field through a unified taxonomy and cross-paper analyses.

Our goal is to offer a consolidated, end-to-end reference for how LLMs are applied to log analysis, what works in practice, and what remains open.
To that end, we make the following contributions:
\begin{itemize}
  \item \textbf{A systematic, task-driven mapping of LLM4Log research (2020--2025).} We provide a curated and categorized corpus, summarize the temporal and task distribution, and characterize where the community concentrates effort and where gaps remain.
  \item \textbf{A pipeline-level review of log-centric LLM methods.} We review and compare LLM-based approaches for (i) logging statement generation and maintenance, (ii) log parsing, and (iii) downstream tasks (anomaly detection, failure prediction, root cause analysis, and log summarization), highlighting reusable design patterns and recurring trade-offs across the log-analysis workflow.
  \item \textbf{Cross-cutting insights, evaluation caveats, and future directions for LLM4Log.} We distill lessons about evaluation practices, robustness under drift and long-tail events, grounding and faithfulness for operator-facing outputs, privacy/security constraints, deployment practicality, and reproducibility, and we summarize these observations in a dedicated cross-cutting section that outlines priorities for reliable real-world adoption.
\end{itemize}

\noindent\textbf{Paper organization.}
The rest of this paper is organized as follows.
Section~\ref{sec:survey_methodology} describes our systematic survey methodology.
Section~\ref{sec:llm_background} provides LLM background and a taxonomy of adaptation/enhancement paradigms for log analysis.
Section~\ref{sec:Logging} reviews LLM-based logging statement generation and maintenance.
Section~\ref{sec:Log_parsing} surveys LLM-based log parsing.
Section~\ref{sec:Log_analysis} covers downstream log analysis tasks, including anomaly detection, failure prediction, root cause analysis, and log summarization.
Section~\ref{sec:cross_cutting} synthesizes recurring insights, open challenges, and future directions across tasks.
Finally, Section~\ref{sec:conclusion} concludes the survey.

%% file: Survey_methodology.tex
\section{Survey Methodology}
\label{sec:survey_methodology}

To build the literature corpus for this survey, we conducted a structured search for publications on \emph{LLM-based log analysis}.
We used a flagship-venue-first search strategy. We started from major software engineering venues to build a focused initial paper set. We recognize that this choice may under-represent some interdisciplinary or industry-facing work.
We therefore complemented the initial venue-based search with backward and forward snowballing, and with digital-library and academic-search cross-checking to broaden coverage.
The literature collection was completed in November 2025. The complete paper list and corpus metadata are available in our public repository: \url{https://github.com/zeyang919/LLM4Log}.

\begin{figure*}[h]
\centering
\footnotesize
\begin{tikzpicture}[
    node distance=0.35cm and 0.35cm,
    box/.style={
        rectangle,
        rounded corners=2pt,
        draw=black!70,
        thick,
        fill=gray!5,
        text width=0.19\textwidth,
        minimum height=2.8cm,
        align=center,
        inner sep=5pt
    },
    finalbox/.style={
        rectangle,
        rounded corners=2pt,
        draw=blue!50!black,
        thick,
        fill=blue!5,
        text width=0.13\textwidth,
        minimum height=2.8cm,
        align=center,
        inner sep=5pt
    },
    arrow/.style={
        -{Latex[length=2.2mm]},
        thick,
        draw=black!70
    }
]

\node[box] (s1) {
    \textbf{Stage 1}\\[2pt]
    \textbf{Flagship-venue-first keyword search}\\[8pt]
    Searched flagship software engineering venues\\[10pt]
    \textbf{52 unique papers}
};

\node[box, right=of s1] (s2) {
    \textbf{Stage 2}\\[2pt]
    \textbf{Backward and forward snowballing}\\[8pt]
    Papers connected to the Stage-1 seed set through\\
    citation/reference links\\[10pt]
    \textbf{70 additional unique papers}\\[4pt]
};

\node[box, right=of s2] (s3) {
    \textbf{Stage 3}\\[2pt]
    \textbf{Digital-library and academic-search cross-check}\\[8pt]
    Remaining papers identified through\\
    database and search cross-checking\\[10pt]
    \textbf{23 additional unique papers}
};

\node[finalbox, right=of s3] (final) {
    \textbf{Final corpus}\\[10pt]
    \textbf{145 unique papers}
};

\draw[arrow] (s1.east) -- (s2.west);
\draw[arrow] (s2.east) -- (s3.west);
\draw[arrow] (s3.east) -- (final.west);

\end{tikzpicture}
\caption{Contribution of the three literature-collection stages to the final corpus of 145 unique papers. The values report the number of papers added at each stage.}
\label{fig:literature_collection_workflow}
\end{figure*}

\subsection{Literature collection protocol}
\label{subsec:literature_collection_protocol}

\noindent\textbf{Stage 1: flagship-venue-first keyword search.}
We began with keyword-based searches in the following software engineering venues, which constituted our Stage-1 flagship-venue-first search set: ICSE, FSE, ASE, ISSTA, ICST, ICSME, ISSRE, ESEM, ICPC, TSE, TOSEM, EMSE, JSS, and IST as shown in Table~\ref{tab:stage1_venues}.
This stage yielded 52 unique papers, as summarized in Figure~\ref{fig:literature_collection_workflow}.
Table~\ref{tab:stage1_venues} summarizes the venues covered in the Stage-1 search.

\begin{table}[h]
  \centering
  \small
  \caption{Stage-1 search venues and year range.}
  \label{tab:stage1_venues}
  \begin{tabular}{p{1.4cm} p{4.6cm} p{1.8cm}}
    \toprule
    Type & Venues & Years \\
    \midrule
    Conference & ICSE, FSE, ASE, ISSTA, ICST, ICSME, ISSRE, ESEM, ICPC & 2020--2025 \\
    Journal & TSE, TOSEM, EMSE, JSS, IST & 2020--2025 \\
    \bottomrule
  \end{tabular}
\end{table}

We used combinations of \emph{log-related} and \emph{LLM-related} keywords.
The log-related keywords include terms such as \emph{``log'', ``system log'', ``runtime log'', ``log analysis'', ``log parsing'', ``log anomaly detection'', ``failure prediction'', ``root cause analysis'', ``incident diagnosis'', ``log summarization''},
the LLM-related keywords include terms such as \emph{``large language model'', ``LLM'', ``Transformer'', ``BERT'', ``GPT'', ``ChatGPT'', ``in-context learning'', ``prompting'', ``retrieval-augmented generation'', and ``agent''}.
We iteratively refined the keyword set as we encountered task-specific terminology and community-preferred phrasing (e.g., operational diagnosis and troubleshooting terms).

\noindent\textbf{Stage 2: snowballing to extend coverage.}
From the seed set obtained in Stage 1, we performed backward snowballing by inspecting references and forward snowballing by checking citations to discover additional relevant papers that may not be easily retrieved via keyword queries, such as papers using non-standard task names or papers whose titles and abstracts emphasize systems and operations without using ``log.''
We also tracked community resources that aggregate log-analysis and LLM-for-operations literature, such as benchmarks, dataset/toolkit papers, and curated lists, as auxiliary entry points.
This stage contributed 70 additional unique papers (Figure~\ref{fig:literature_collection_workflow}).

\noindent\textbf{Stage 3: digital-library cross-check.}
To reduce omission risk and broaden coverage beyond the Stage-1 venue set, we cross-checked our corpus using major digital libraries and preprint servers (i.e., IEEE Xplore, ACM Digital Library, SpringerLink, Elsevier/ScienceDirect, Wiley Online Library, arXiv, and OpenReview), and used Google Scholar for additional coverage checks and citation tracing.
This stage contributed the remaining 23 unique papers.
Figure~\ref{fig:literature_collection_workflow} summarizes how the three stages contributed to the final corpus of 145 unique papers.

\noindent\textbf{Scope control and manual screening.}
Keyword searches inevitably return false positives due to term ambiguity and task overlap.
For example, some papers include both \emph{``root cause analysis''} and \emph{``LLM''} keywords, but evaluate only metrics or traces, while some anomaly-detection papers use network flows, sensor telemetry, or general text without runtime logs.
Similarly, the term ``log'' can refer to non-runtime logs (e.g., query logs, change logs, process mining event logs) or even mathematical ``log'' usage.
Candidate papers were manually screened by examining their titles, abstracts, and relevant method and evaluation sections.
Inclusion was decided at the paper level using the criteria below.
We consolidated repeated occurrences of the same paper across task assignments and collection stages by title and stable identifier.
For downstream tasks, we distinguish three cases and apply them consistently across all task sections.
Studies that explicitly evaluate software or system runtime logs as the main input are within the log-specific scope.
Multimodal studies are also within scope when runtime logs are explicitly evaluated together with at least one other operational data source, such as metrics, traces, code, tickets, or topology.
When logs are optional or used only in some experimental settings, only results that explicitly involve runtime logs are used to support log-specific findings.
If those results cannot be separated, the study may be discussed as a related comparison, but it is not used as evidence about log analysis.

Runtime logs in this review include application, system, security/audit, shell, build, test, and console event records produced during software or system operation.
Host audit and provenance records qualify when they record process actions or system calls.
Metrics, distributed traces, network packets or flows, sensor telemetry, tickets, and incident summaries do not qualify by themselves.
A derived artifact qualifies only when the study documents that it was constructed directly from runtime logs.

\noindent\textbf{Inclusion and exclusion criteria.}
A publication contributes log-specific evidence if it satisfies all of the following:
(1) it targets one or more tasks represented in our taxonomy (e.g., logging, parsing/representation, anomaly/failure detection, failure prediction, diagnosis/RCA, or summarization).
(2) It uses a Transformer-based pretrained language model (encoder-only, encoder--decoder, or decoder-only) via prompting/ICL, retrieval grounding, fine-tuning/post-training, tool augmentation/agents, or representation learning.
(3) It reports an empirical evaluation, ablation/analysis, or a concrete case study in which logging statements are the primary technical output or runtime logs are an explicit input to the method or benchmark, or it presents a concrete technical design directly focused on logging.
Concrete vision or design records are described as such and are not used as empirical evidence.
Related comparisons are identified in the discussion and are not treated as evidence for log-specific findings.
We exclude work whose task is outside our taxonomy.

\noindent\textbf{Data extraction protocol.}
For each paper, we extracted structured information to support later taxonomy construction and cross-paper comparison.
When a paper covered several tasks, we created a task-paper record for each assignment and extracted fields relevant to that task.
Across tasks, we recorded: (i) target task(s), (ii) dataset type and whether logs were used alone or with auxiliary artifacts (e.g., metrics/traces/tickets/runbooks/code/config), and (iii) evaluation setup.
In addition, for each task, we explicitly extracted task-relevant technical details, such as:
\emph{baseline LLM(s)}, \emph{LLM enhancement/adaptation paradigm(s)} (e.g., ICL, retrieval grounding/RAG, fine-tuning, tool/agent augmentation, verification), and \emph{evaluation metric(s)}.
The complete paper list and extracted metadata are available in our public repository.

\begin{figure}[t]
  \centering
  \includegraphics[width=0.65\columnwidth]{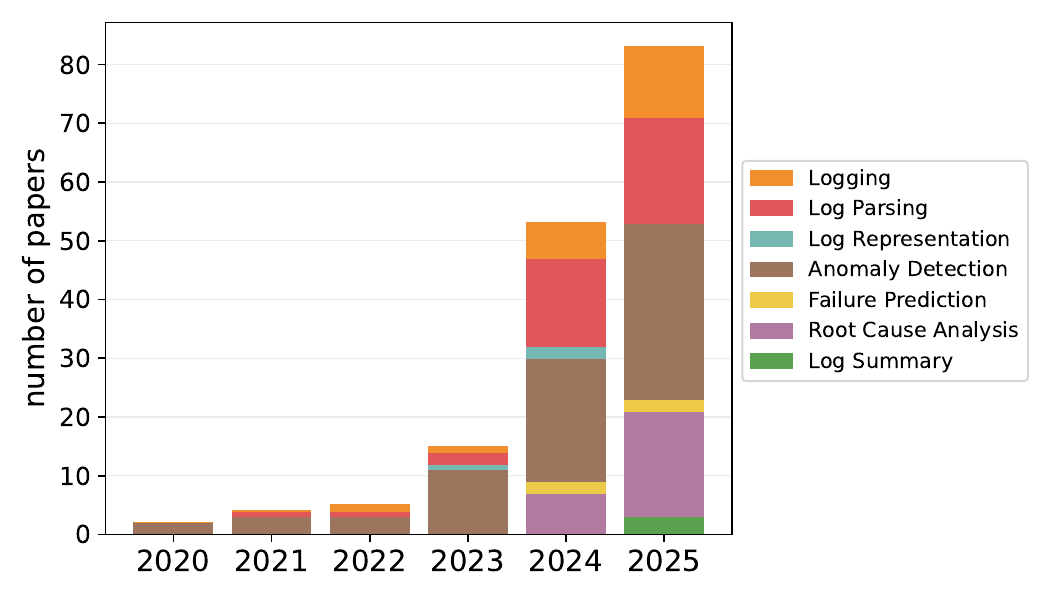}
  \includegraphics[width=0.3\columnwidth]{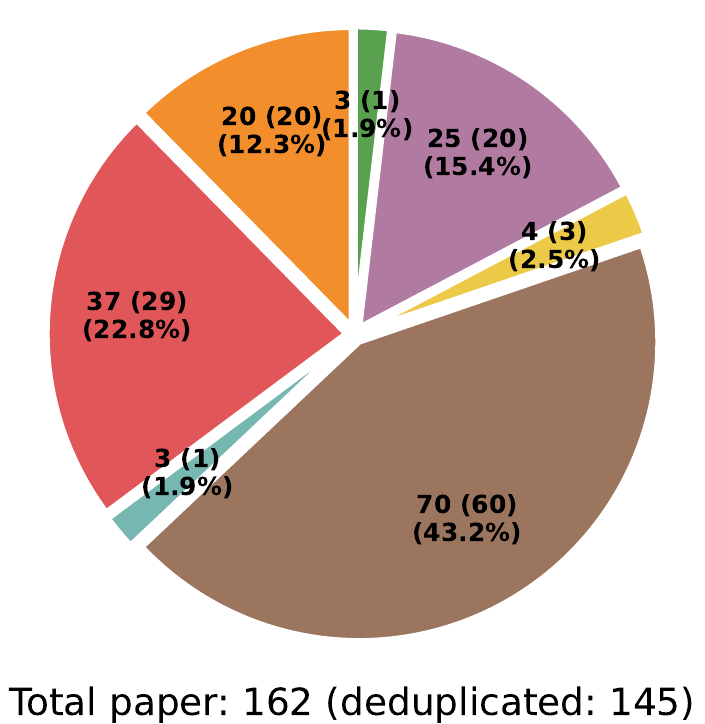}
  \caption{Temporal and task distribution of the 162 task-paper records. Labels use $n$ ($u$), where $n$ is the number of task-paper records and $u$ is the number of papers assigned only to that task. The figure also reports the deduplicated total of 145 unique papers.
  }
  \label{fig:paper_summary}
\end{figure}

\Needspace{11\baselineskip}
\subsection{Corpus summary and categorization.}
Based on our task-driven screening, we reviewed 162 task-paper records spanning seven log-analysis tasks, published between 2020 and 2025.
The literature search was completed in November 2025.
Deduplicating these records by paper identity yields 145 unique papers.
He et al.~\cite{he2021survey} reported 158 log-analysis papers from 1997 to 2020 under a different collection protocol.
Although the totals are not directly comparable, the 145 papers collected here over 2020--2025 show the rapid growth of LLM-based log analysis.
Figure~\ref{fig:paper_summary} shows the temporal trend and task composition of the full reviewed set.

\Needspace{18\baselineskip}
\begin{wraptable}[18]{r}{0.4\columnwidth}
  \centering
  \small
  \caption{Top venues by number of papers in our LLM4Log corpus.}
  \label{tab:top_venues}
  \begin{tabular}{l r}
    \toprule
    Venue & \#Papers \\
    \midrule
    arXiv & 29 \\
    ICSE & 11 \\
    ASE & 7 \\
    ISSRE & 7 \\
    FSE & 6 \\
    CIKM & 5 \\
    ICLR & 5 \\
    EMSE & 4 \\
    TOSEM & 4 \\
    ICPC & 3 \\
    \bottomrule
  \end{tabular}
  \begin{tablenotes}
    \footnotesize
  \item \textbf{Note:} We report the top 10 venues in the paper. The full list is available in our online repository.
  \end{tablenotes}
\end{wraptable}

The task distribution (Fig.~\ref{fig:paper_summary}, right) is highly skewed.
Anomaly Detection and Log Parsing together account for roughly \emph{two-thirds} of all task-paper records, reflecting the early use of LLMs to identify suspicious behavior in noisy, drifting logs and to structure raw logs for later analysis.
At the same time, the corpus also shows a clear expansion toward more downstream, operator-facing tasks.
In particular, Root Cause Analysis forms a sizable slice and grows rapidly in the most recent years, indicating a shift from ``detecting something is wrong'' toward ``explaining why and what to do,'' often requiring richer reasoning and evidence integration.

Temporally, the corpus spans 2020--2025 and exhibits a pronounced surge in recent years (Fig.~\ref{fig:paper_summary}, left).
Only a small number of papers appeared before 2023, followed by a sharp jump in 2024 and an even larger volume in 2025.
Beyond the overall growth, the per-task evolution is also informative: anomaly detection appears earliest and grows steadily, while several downstream tasks emerge later and accelerate quickly (e.g., RCA and log summarization are largely concentrated in the most recent period).
This pattern suggests that the community first adopted LLMs where they could be plugged into existing pipelines (detection/parsing) and then progressively pushed them toward higher-level operational assistance as models and prompting/retrieval techniques matured.

Finally, the venue landscape indicates both broad dissemination and strong SE anchoring.
As summarized in Table~\ref{tab:top_venues}, a noticeable fraction of the corpus appears as preprints (reflecting the fast-moving nature of LLM research), while many peer-reviewed papers are mainly published in software engineering venues.
Interestingly, the venue concentration differs by task: papers on anomaly detection are much more widely distributed across research communities and venues (e.g., systems, security, networking, data mining, and AI), whereas topics closer to software engineering practice (e.g., logging and some parsing work) are more likely to appear in SE-focused outlets.
This cross-venue spread reinforces that LLM4Log is inherently interdisciplinary: it inherits operational problems from SE while drawing methods and evaluation norms from multiple adjacent fields.

\subsection{Relation to existing surveys.}
In recent years, several surveys and reviews have summarized progress in automated log analysis and related LLM-based operational analytics.
He et al.~\cite{he2021survey} offered a broad survey of automated log analysis for reliability engineering, summarizing established pipelines spanning logging practice, compression, parsing, anomaly detection, failure prediction, and diagnosis, and largely reflecting the pre-instruction-tuned-LLM era.
As LLMs entered log analytics, more focused surveys emerged. Beck et al.~\cite{beck2025system} concentrated on \emph{system log parsing with large language models}, emphasizing how LLMs convert raw logs into structured templates and fields and how such parsers are evaluated. Akhtar et al.~\cite{akhtar2025llm} reviewed \emph{LLM-based event log analysis} in a broader sense, motivated primarily by the needs of security professionals and covering event logs from multiple computing contexts, including system, application, and network logs. Their survey mainly summarizes what LLM techniques have been applied to event logs, while our work uses software runtime logs as the organizing lens for task structure, evaluation practice, and open challenges.

Several adjacent surveys also provide useful comparison points without being log-centric in the same way as our work. Remil et al.~\cite{remil2024aiops} reviewed AIOps solutions for incident management across tasks, data sources, and technical approaches. Zhou and Fokaefs~\cite{zhou2024incident} surveyed AI assistants for the incident lifecycle in microservice environments, considering observability data such as logs, traces, and metrics. Wang and Qi~\cite{wang2024rca} surveyed root cause analysis in (micro)services across logs, traces, metrics, and multimodal data. From the operational analytics side, Zhang et al.~\cite{zhang2025survey} discussed AIOps in the LLM era at a broader reliability-engineering level, where logs are one type of operational data among metrics, traces, alerts, incidents, and other signals.

Overall, our survey complements these efforts by using software and system runtime logs as the organizing evidence across the end-to-end workflow.
For downstream tasks, we apply the same three-case policy: studies are within the log-specific scope when runtime logs are the main evaluated input or an explicit part of multimodal evidence. When log use is optional, only separable results that explicitly involve runtime logs support log-specific findings. Studies without separable log-specific results may be discussed as related comparisons, but they do not support findings about log analysis.
As summarized in Table~\ref{tab:survey_comparison}, the survey spans logging, parsing, and downstream analysis while examining how LLM-based methods interact with deployment constraints and evidence quality.

\Needspace{16\baselineskip}
\begin{table*}[t]
  \centering
  \small
  \caption{Positioning of this survey relative to representative prior surveys.}
  \label{tab:survey_comparison}
  \begin{tabular}{p{3.0cm} p{3.8cm} p{0.5cm} p{6.8cm}}
    \toprule
    Survey & Focus & LLM & Difference from our survey \\
    \midrule
    He et al.~\cite{he2021survey} & Broad log analysis before the recent LLM wave & No & Updates the field for the LLM era with a dedicated log-centric focus \\
    Beck et al.~\cite{beck2025system} & LLM-based system log parsing & Yes & Extends beyond parsing to the full log-analysis workflow \\
    Akhtar et al.~\cite{akhtar2025llm} & LLM-based event-log analysis across security and computing contexts & Yes & Mainly summarizes what LLM techniques exist for event-log analysis. Our survey focuses on how LLM4Log is structured, evaluated, and where open challenges remain \\
    Remil et al.~\cite{remil2024aiops} & AIOps incident-management review & No & Broader incident-management scope without a dedicated LLM-based log-centric analysis \\
    Zhou and Fokaefs~\cite{zhou2024incident} & AI assistants for incident lifecycle in microservices & No & Incident-lifecycle support across logs, traces, and metrics outside a log-centric workflow \\
    Wang and Qi~\cite{wang2024rca} & RCA survey in (micro)services & No & Focuses on one downstream task and broader observability signals, not the full LLM4Log workflow \\
    Zhang et al.~\cite{zhang2025survey} & LLM-enabled AIOps and reliability engineering & Yes & Covers broader LLM4AIOps workflows across logs, metrics, traces, alerts, and incidents. Our survey focuses on how LLM methods are adapted and evaluated when runtime logs are the central evidence source \\
    \bottomrule
  \end{tabular}
\end{table*}

\subsection{Threats to validity.}
Our survey collection methodology faces several limitations.
First, the flagship-venue-first search strategy improves precision and anchors the survey in core software engineering venues, but it may still miss some interdisciplinary, industry-facing, or rapidly emerging work despite the subsequent snowballing and digital-library cross-checking.
Second, the search process depends on keyword design, so relevant papers that use uncommon terminology may be harder to retrieve in the initial stages.
Third, deciding whether runtime logs are explicitly evaluated inputs still requires manual judgment because terms such as ``log,'' ``event,'' ``trace,'' and ``incident record'' are not used consistently across papers.
We reduce this ambiguity by applying the definition above and reporting evaluated input types.
Because our retrieval strategy was centered on log-related terms, the related comparisons provide context but not a complete review of non-log or modality-agnostic LLM4AIOps research.

Cross-paper evaluation remains a threat to validity. The studies covered in this survey differ in their datasets, label definitions, model families, prompts, decoding settings, and evaluation metrics. In addition, many papers do not report enough prompt or inference details to support exact reproduction. For LLM-based methods, results may also vary across runs because of stochastic decoding, model updates, retrieval choices, or context selection. These differences make it difficult to compare reported performance numbers directly across papers.
For this reason, our main conclusions focus on recurring design patterns, common evaluation gaps, and practical constraints in LLM4Log research. We treat claims about the best-performing method as exploratory.

%% file: background.tex
\section{LLM Background \& Taxonomy for Log Analysis}
\label{sec:llm_background}

LLMs have become a practical tool for log-centric software operations because logs are semi-structured, drift over time, preserve execution chronology, and are typically interpreted alongside heterogeneous artifacts (tickets, dashboards, traces, configurations, and code)~\cite{llmparser2024icse,mastropaolo2022using,xu2023logppt,Guo2021LogBERT}.
This section explains why logs form a distinct setting for LLM-based analysis, how major model families map to log-analysis workloads, and how adaptation paradigms interact with operational constraints such as drift, grounding, privacy, latency, and cost.

\subsection{Why LLMs are a Good Fit for Log Tasks}
\label{subsec:llm_why_logs}

\phead{Logs as semi-structured text that bridges natural language and code.}
A key reason LLMs fit log analysis is that log messages sit between natural language and code.
A typical log line mixes free-form descriptions (symptoms, actions, causal hints) with code-like structure and identifiers (function and class names, stack frames, paths, IPs, config keys, request IDs, error codes)~\cite{adapting2025arxiv,jiang2024prelog}.
This hybrid format aligns with two core LLM strengths: natural-language understanding and code-aware pattern modeling~\cite{devlin2019bert,mastropaolo2022using}.
At the same time, log analysis is not just generic text understanding: the boundary between stable templates and variable values is often ambiguous, and the same token may be a symptom, parameter, component name, or incident clue depending on context.
As a result, LLMs can capture intent and semantics that are often discarded when logs are reduced to templates/event IDs, and can still remain robust when token vocabularies evolve (e.g., new components, identifiers, or new error strings)~\cite{llmparser2024icse,li2024lilac}.

\phead{Latent semantics beyond templates and surface patterns.}
Although many classic pipelines reduce logs to templates/event IDs, real log messages encode operational meaning:
what operation was attempted, what failed, which component is implicated, how severe it is, and what the system was doing right before/after~\cite{jiang2024prelog,xu2023logppt}.
These signals are frequently expressed implicitly via linguistic cues (error idioms, temporal markers, causal connectors, negation), code-like cues (exception types, API names, call sites, module prefixes), and temporal ordering across neighboring events~\cite{Guo2021LogBERT,llmparser2024icse}.
This makes logs different from many standalone text artifacts: they often carry a symptom chronology that operators use to reconstruct incidents.
LLMs can exploit these cues to support interpretation, explanation, and cross-source integration, especially when purely symbolic representations are brittle or incomplete~\cite{owl2024iclr,Karlsen2024BenchmarkingLLMsJNSM,ma2024luk}.

\phead{Robustness under drift, partial observability, and long-tail behaviors.}
Logs evolve with versions, configuration changes, feature flags, and deployment environments. Meanwhile, rare events are common in operations and often carry disproportionate diagnostic value, even when they appear too infrequently for stable supervised learning~\cite{llmparser2024icse,li2024lilac}.
Compared with fixed-vocabulary or rule-based parser-dependent pipelines, LLM-based representations and prompting enable soft matching and semantic generalization, which helps when (i) rule-based parsers fail or drift, (ii) labeled anomalies/failures are scarce, or (iii) the same underlying issue is described with heterogeneous phrasing across services and versions~\cite{xu2024divlog,comparative2024esem,llmparser2024icse}.

\phead{Unifying heterogeneous evidence into a single reasoning space.}
In production triage, the relevant evidence rarely lives in logs alone~\cite{owl2024iclr,Cui2024LogEval}.
Engineers correlate logs with traces/metrics, tickets, runbooks, configurations, and sometimes code diffs~\cite{owl2024iclr,Cui2024LogEval}.
LLMs can condition on mixed inputs (log snippets, retrieved historical cases, metric summaries, runbook steps, code/config fragments) and produce operator-oriented outputs such as hypotheses, ranked causes, incident narratives, and action candidates~\cite{owl2024iclr,ma2024luk,Karlsen2024BenchmarkingLLMsJNSM}.
This ability to work across heterogeneous artifacts matches the reality of AIOps workflows, but a log-centric view remains necessary because logs often provide the chronological symptom evidence around which other signals are interpreted~\cite{owl2024iclr,Karlsen2024BenchmarkingLLMsJNSM}.

\phead{Human-facing outputs and decision support.}
Many operational tasks require a narrative (what happened, what changed, why likely) and not just a score/label~\cite{owl2024iclr,Karlsen2024BenchmarkingLLMsJNSM}.
LLMs can highlight evidence, translate technical fragments into concise explanations, and generate structured artifacts for handoff (e.g., timelines, suspected components, extracted entities)~\cite{ma2024luk,ma2025logreasoner,Cui2024LogEval}.
However, free-form generation increases the risk of hallucinated rationales or overconfident claims~\cite{Karlsen2024BenchmarkingLLMsJNSM,Cui2024LogEval}.
In practice, LLMs are most reliable when outputs are constrained and augmented (e.g., by quoting evidence spans, using structured extraction, or tying claims to retrieved/tool-produced facts) rather than relying on unconstrained storytelling~\cite{ma2024luk,ma2025logreasoner,owl2024iclr}.

\subsection{Evolution of LLMs: Architectures and Training Stages}
\label{subsec:llm_evolution}

\phead{Transformer backbones for long-context log modeling.}
Most LLM-based log-analysis methods build on Transformer architectures~\cite{vaswani2017attention}, whose self-attention mechanism supports scalable contextual modeling across long input windows.
This matters for logs because useful cues can be far apart (e.g., an early warning followed by a later error burst), and multi-line context often changes the meaning of individual events~\cite{vaswani2017attention}.

\phead{Encoder-only models: masked language modeling and contextual embeddings.}
Encoder-only pretraining with masked language modeling (MLM), popularized by BERT~\cite{devlin2019bert} and its variants (e.g., RoBERTa~\cite{liu2019roberta}, ELECTRA~\cite{clark2020electra}), learns bidirectional contextual representations.
Compared with decoder-only chat-style LLMs, encoder-only models typically have fewer parameters and lower inference cost, and some log-analysis papers therefore refer to them as \emph{pretrained language models (PLMs)}, not LLMs. However, since they are still substantially larger than traditional log models and have a large body of log-centric research, we treat them as part of the broader LLM family in this survey~\cite{qiu2020ptmsurvey}.
In log analytics, encoder-only LMs are most useful as efficient representation and scoring backbones~\cite{devlin2019bert,Guo2021LogBERT}.
Their pretraining objectives can support semi- or unsupervised anomaly signals, while their embeddings support retrieval, clustering, and lightweight downstream heads for anomaly/failure classification when some supervision exists~\cite{devlin2019bert,clark2020electra,Guo2021LogBERT,reimers2019sentencebert,karpukhin2020dpr}.

\phead{Encoder--decoder models: conditional generation and structured transformation.}
Encoder--decoder Transformers generalize from representation learning to conditional generation: an encoder ingests the input sequence and a decoder produces an output sequence~\cite{vaswani2017attention,sutskever2014sequence}.
Representative families include T5~\cite{raffel2020exploring} and BART~\cite{lewis2020bart}, which are typically trained with denoising or text-to-text objectives~\cite{raffel2020exploring,lewis2020bart,vincent2008denoising}.
For logs, this paradigm is most relevant when the task transforms one textual form into another: incident summarization, noisy-message normalization, structured extraction, or template-like rewriting conditioned on surrounding context~\cite{lewis2020bart,raffel2020exploring,see2017get}.

\phead{Decoder-only models: autoregressive pretraining and general-purpose assistance.}
Decoder-only autoregressive LMs dominate interactive settings because they are flexible under prompting and can be scaled to strong instruction-following behavior~\cite{brown2020language,kaplan2020scaling}.
Representative examples include the GPT family~\cite{brown2020language} and open-weight decoder models such as LLaMA-style backbones~\cite{touvron2023llama}.
For log analysis, they are attractive when workflows require open-ended reasoning, hypothesis generation, retrieval-augmented triage, remediation suggestions, or multi-turn operator interaction, especially when labeled data is scarce~\cite{wei2022emergent,ouyang2022instructgpt}.
The trade-off is higher inference cost and latency, which motivates selective escalation and hybrid designs discussed later~\cite{shoeybi2019megatronlm}.

\phead{Key capabilities that matter for log tasks.}
Across architectures, several capabilities are particularly relevant to log analysis~\cite{bommasani2021foundation,gpt4report2023,lewis2020rag,wei2022cot,yao2022react,ouyang2022instructgpt}:
\textbf{(1) semantic normalization and abstraction}---mapping heterogeneous phrasings to consistent meanings and grouping variants of the same issue.
\textbf{(2) contextual disambiguation}---interpreting the same token differently depending on surrounding components, timestamps, and execution context (crucial under drift).
\textbf{(3) evidence integration over long contexts}---integrating signals across many lines and multiple sources (logs, traces, metrics, tickets), typically mediated by windowing or retrieval in practice.
\textbf{(4) controllable outputs}---producing operator-facing artifacts (ranked lists, extracted entities, timelines) instead of only free-form text.
\textbf{(5) reasoning and decomposition}---breaking complex tasks into steps (form hypotheses, retrieve context, verify, refine), which underpins agentic log analysis pipelines.
\textbf{(6) software and code awareness}---understanding stack traces, API names, configuration keys, and common failure modes, enabling more actionable interpretation than purely surface-level text matching~\cite{chen2021codex}.

\subsection{LLM Capability and Specialization Taxonomy for Log Analytics}
\label{subsec:llm_taxonomy}

LLMs used in log analysis can be grouped by specialization level, which often determines where they fit in the log pipeline~\cite{bommasani2021foundation}.

\phead{General-purpose LLMs} are trained broadly on large-scale, diverse text corpora and excel at natural-language interpretation, summarization, and interactive assistance~\cite{brown2020language,gpt4report2023}.
They are useful for prompt-based log understanding, RCA narratives, and incident reporting, but usually need retrieved context to handle organization-specific terminology, component names, and operational conventions~\cite{gpt4report2023,ouyang2022instructgpt,gururangan2020dontstoppretraining,lewis2020rag}.

\phead{Code-oriented LLMs} are trained or fine-tuned heavily on code and technical corpora, improving performance on stack traces, error messages, configuration files, and program-context reasoning~\cite{chen2021codex,feng2020codebert,wang2021codet5}.
They are a better fit when log interpretation depends on linking symptoms to code/config, call stacks, APIs, or remediation commands~\cite{chen2021codex}.

\phead{Domain-tailored LLMs} adapt a backbone to a specific operational domain (e.g., AIOps for cloud platforms, network diagnostics, ICS security) via continued pretraining, instruction tuning, or explicit knowledge integration~\cite{gururangan2020dontstoppretraining,ouyang2022instructgpt}.
They are most valuable when logs contain specialized vocabularies, recurring incident patterns, and internal operational procedures that general-purpose models cannot infer reliably~\cite{gururangan2020dontstoppretraining}.

\subsection{Adaptation and Enhancement Paradigms}
\label{subsec:llm_enhancement}

Off-the-shelf LLMs are often not trained \emph{specifically} for log analysis: they rarely see an organization’s proprietary log vocabulary, evolving component taxonomy, or operational ground-truth data (incidents, runbooks, and on-call decisions)~\cite{bommasani2021foundation,gururangan2020dontstoppretraining}.
Therefore, LLM-based log analysis approaches typically improve reliability by controlling \emph{what} the model sees (context and retrieved evidence) and \emph{how} it decides (prompting, reasoning scaffolds, tuning, tools, and verification)~\cite{lewis2020rag,yao2022react,wei2022cot,ouyang2022instructgpt}.
Below, we summarize the most common paradigms in terms of the log-specific trade-offs they address: label scarcity, drift, grounding, controllability, inference cost, and deployment practicality~\cite{bommasani2021foundation}.

\phead{Prompting and in-context learning (ICL), including retrieval grounding.}
ICL treats the LLM as a training-free learner: prompts provide task instructions, output constraints, and a small set of demonstrations~\cite{brown2020language,liu2021pretrainpromptpredict,min2022rethinkingicl}.
In log settings, effective ICL is rarely a single ``classify this log'' instruction. It typically includes:
(i) \textbf{input normalization} (variable masking, deduplication, component/time grouping),
(ii) \textbf{bounded evidence} (top-$k$ suspicious lines/windows rather than full streams),
(iii) \textbf{schema-constrained outputs} (labels, ranked candidates, JSON fields),
and (iv) \textbf{system-specific exemplars} (a few representative normal/failure patterns from the same environment)~\cite{liu2021pretrainpromptpredict,min2022rethinkingicl}.

A common extension is \textbf{retrieval-augmented generation (RAG)}, which retrieves relevant context (e.g., similar historical windows/incidents, runbook passages, known signatures, or indexed summaries) and injects it into the prompt as additional demonstrations or evidence~\cite{lewis2020rag}.
In many log-analysis pipelines, RAG can be viewed as an \emph{ICL enabler} rather than a separate paradigm: retrieval supplies high-quality in-context examples and system knowledge at inference time, without changing model weights~\cite{lewis2020rag}.
This ``retrieve $\rightarrow$ prompt $\rightarrow$ decide'' pattern improves both \emph{grounding} (anchoring outputs to concrete evidence) and \emph{adaptation} (bringing system-specific knowledge under drift)~\cite{lewis2020rag,gururangan2020dontstoppretraining}.

ICL remains attractive for rapid deployment and heterogeneous tasks, but it is sensitive to prompt quality, context selection, and distribution shift. As a result, many practical approaches combine RAG-augmented ICL with lightweight filtering (to reduce prompt noise) and downstream verification (to prevent overconfident hallucinations)~\cite{liu2021pretrainpromptpredict,min2022rethinkingicl}.
\phead{Reasoning scaffolds: CoT, ToT, and structured decomposition.}
Many log tasks (especially diagnosis and remediation) benefit from explicit decomposition rather than one-shot generation~\cite{wei2022cot,zhou2022leasttomost}.
\textbf{Chain-of-Thought (CoT)} prompting encourages stepwise reasoning (e.g., identify key symptoms $\rightarrow$ map to components $\rightarrow$ hypothesize causes $\rightarrow$ justify with evidence)~\cite{wei2022cot}.
In log settings, CoT is most useful when intermediate artifacts are \emph{checkable} and \emph{grounded} (e.g., extracted entities, cited log lines, matched templates), not when it produces purely speculative narratives~\cite{wei2022cot,wang2022selfconsistency}.
\textbf{Tree-of-Thought (ToT)} generalizes CoT by exploring multiple candidate hypotheses/paths, scoring or pruning them, and selecting the best~\cite{yao2023tot,wang2022selfconsistency}.
This matches operational diagnosis where multiple plausible causes exist. ToT-style search can be implemented via self-consistency sampling, explicit branching (e.g., per service/component), and re-ranking using retrieved evidence or tool outputs~\cite{yao2023tot,wang2022selfconsistency}.
The practical trade-off is cost: ToT multiplies calls/tokens, so it is often reserved for high-severity complex incidents, and paired with retrieval and/or a smaller model to narrow the search space~\cite{yao2023tot}.
More generally, many approaches adopt \textbf{structured decomposition} without explicitly naming CoT/ToT by splitting the workflow into stages such as symptom extraction, candidate generation, evidence retrieval, and final decision~\cite{zhou2022leasttomost,yao2022react}.

\phead{Fine-tuning and post-training for log tasks.}
When data is available, training often yields better stability than pure prompting~\cite{gururangan2020dontstoppretraining}.
It is useful to distinguish:
(i) \textbf{continued pretraining} (domain-adaptive pretraining on large unlabeled log corpora) to improve coverage of log vocabulary and style,
(ii) \textbf{supervised fine-tuning (SFT)} on labeled log tasks,
(iii) \textbf{parameter-efficient tuning (PEFT)} (e.g., LoRA/adapters) to reduce cost and enable more frequent updates under drift,
and (iv) \textbf{post-training/alignment} (instruction tuning and preference optimization) to improve instruction following, format adherence, and safety/harmlessness behaviors~\cite{ouyang2022instructgpt,rafailov2023dpo}.
In log analysis, post-training is especially relevant when the model must (a) reliably follow output schemas, (b) avoid leaking sensitive data in summaries, and (c) produce calibrated, non-overconfident explanations (including the ability to abstain)~\cite{ouyang2022instructgpt}.
A common pattern is \textbf{hybrid training}: fine-tune a smaller encoder/LM for scoring or classification while using a larger aligned model for explanation and interactive assistance~\cite{bommasani2021foundation,gpt4report2023}.
Another pattern is \textbf{task-to-instruction transformation}, where multiple log tasks are cast into a unified instruction format so one model can support classification, extraction, and generation consistently~\cite{ouyang2022instructgpt}.

\phead{Tool augmentation and agentic workflows.}
Agentic log-analysis approaches extend in-context retrieval by letting an LLM iteratively call external tools to gather, filter, and validate evidence~\cite{yao2022react,mialon2023augmentedlms,schick2023toolformer}.
Typical tools include log search/index queries, metric/trace dashboards, ticket/KB/runbook lookup, config/code retrieval, and lightweight parsers or summarizers that pre-compress high-volume inputs~\cite{mialon2023augmentedlms}.
This paradigm is a natural fit for real-world log analysis because a single prompt window is rarely sufficient: relevant signals may be scattered across long time spans, multiple components, and heterogeneous repositories~\cite{mialon2023augmentedlms,yao2022react}.

Agentic workflows commonly follow a loop of:
(i) \textbf{planning} (what evidence to obtain next and which hypotheses to test),
(ii) \textbf{acting} (tool calls for retrieval, slicing, aggregation, or correlation),
(iii) \textbf{observing} (interpreting tool outputs and updating the working context),
and (iv) \textbf{deciding} (producing a final prediction/explanation/summary with evidence pointers)~\cite{yao2022react}.
The benefits are stronger grounding and broader coverage. The costs are latency, more failure modes (tool errors, cascading mistakes), and higher security risks (e.g., prompt injection via retrieved documents)~\cite{mialon2023augmentedlms,schick2023toolformer}.
Accordingly, practical deployments often add guardrails such as allow-listed tools, bounded iterations, schema validation, and conservative fallback behaviors~\cite{mialon2023augmentedlms,shinn2023reflexion}.

\phead{Verification, calibration, and controllability.}
Because log analysis is high-stakes and noisy, many approaches add explicit reliability layers:
\textbf{self-consistency} (sample multiple reasoning paths and aggregate),
\textbf{cross-checking} (verify claims against retrieved lines/windows or tool outputs),
\textbf{schema validation} (force typed/JSON outputs and reject invalid responses),
\textbf{evidence citation} (require each claim to point to a log span or retrieved document),
and \textbf{selective generation} (generate only after candidate filtering by a smaller model)~\cite{wang2022selfconsistency,ji2023hallucination,manakul2023selfcheckgpt,kadavath2022know}.
These paradigms reappear throughout the task sections. Section~\ref{sec:cross_cutting} later compares their trade-offs in terms of supervision needs, inference cost, robustness under drift, controllability, and operational fit.

\subsection{Practical Concerns: Privacy, Security, Compliance, and Efficiency}
\label{subsec:llm_practical}

Logs may contain sensitive identifiers (user IDs, IPs, hostnames), infrastructure topology, security indicators, and occasionally personal data~\cite{nist80092logmgmt,gdpr2016}.
As a result, deploying LLMs for log analysis is often governed as much by \textbf{privacy/security constraints} as by model accuracy, while \textbf{latency and cost} remain first-class operational requirements~\cite{gdpr2016,nist80053r5}.
These concerns are widely recognized in the literature, but rigorous production-scale validation remains limited and uneven across tasks.

\phead{Commercial LLMs vs.\ self-hosted LLMs.}
Commercial LLMs are typically consumed via \emph{remote APIs}, meaning prompts (and any retrieved context inserted into them) must be transmitted to an external model provider~\cite{gdpr2016,nist80053r5}.
This increases \textbf{data exposure risk} and reduces end-to-end controllability: even with contractual terms and vendor assurances, organizations may have limited visibility into provider-side handling (e.g., retention, secondary processing, or cross-region routing) and must treat the prompt as a potentially sensitive payload~\cite{carlini2021extracting,gdpr2016}.
In contrast, self-hosted (open-source) models keep inference \emph{within the organization's trust boundary}, making data handling more controllable: access can be restricted via internal IAM, prompts/outputs can be logged or disabled under local policy, and compliance requirements (e.g., residency) can be enforced operationally~\cite{nist80053r5}.
However, self-hosting shifts responsibility to the operator: the main risks become the local security posture, including securing model-serving endpoints, hardening the retrieval stack, and protecting stored prompts, retrieved documents, and generated outputs~\cite{nist80053r5}.

Tool-augmented and retrieval-based workflows amplify these issues regardless of deployment. Untrusted retrieved text can introduce prompt-injection style instructions, and generated summaries/explanations can inadvertently echo secrets or sensitive identifiers if not constrained~\cite{greshake2023indirectpromptinjection,carlini2021extracting}.

\phead{Common mitigations in log analysis pipelines.}
Typical mitigations include:
(i) \textbf{minimization} (send only the smallest relevant window and avoid full dumps),
(ii) \textbf{redaction/masking} (identifiers, secrets, tokens),
(iii) \textbf{bounded retrieval disclosure} (retrieve on-prem, but pass only small snippets/fields),
(iv) \textbf{RBAC and auditing} (who can query what and how prompts/outputs are safely logged),
and (v) \textbf{policy-constrained outputs} (no secret echoing, structured extraction, and citation requirements)~\cite{gdpr2016,nist80053r5,nist80092logmgmt}.
These controls interact with utility: aggressive redaction may remove key signals, so many approaches combine masking with structured extraction and retrieval to preserve diagnostic content while limiting exposure~\cite{gdpr2016,nist80053r5}.

\phead{Cost drivers and practical optimizations.}
The main cost drivers are context length (tokens), number of model calls (multi-pass/agentic workflows), and retrieval/embedding computation~\cite{dettmers2023qlora,pope2023vllm}.
Common optimizations include windowing and deduplication, template compaction, hierarchical summarization, caching embeddings and intermediate summaries, batching, and \textbf{selective escalation} (invoke a large model only after a cheaper filter flags candidates)~\cite{mialon2023augmentedlms}.
For self-hosted models, quantization and efficient serving reduce inference cost. For API models, prompt compression and retrieval-first designs reduce token budgets~\cite{dettmers2023qlora,pope2023vllm}.

\phead{Hybrid designs as a practical deployment pattern.}
A frequent design is \textbf{small model for scoring/filtering} (ranking suspicious windows, candidate component selection, coarse labeling) plus \textbf{large model for explanation and summarization}~\cite{bommasani2021foundation,gpt4report2023}.
This keeps expensive reasoning focused on a small subset of high-value context, reduces latency, and yields operator-facing artifacts (summaries, rationales, action candidates) without incurring LLM costs on every log line~\cite{mialon2023augmentedlms,pope2023vllm}.
In practice, such hybrids also simplify governance: only the most capable (and often most sensitive/costly) components are exposed to the most sensitive or costly data~\cite{gdpr2016,nist80053r5}.

%% file: Logging.tex
\section{Logging Statement Generation and Maintenance}
\label{sec:Logging}
Logging statement generation studies how to automatically insert and update log statements in source code so that systems emit informative runtime records for debugging and monitoring.
In practice, insufficient logging can hinder incident triage and reduce observability, whereas excessive or inconsistent logging introduces noise, performance overhead, and scalability concerns.
Hence, one active line of research is automated logging statement generation, which aims to recommend or synthesize high-quality log statements so that developers can obtain effective logs with less manual effort.

\subsection{Task Definition}
Logging statement generation is the task of inserting and updating log statements into source code so that, during execution, the system emits textual records (logs) describing internal states and runtime events.
Given a code context (e.g., a method, a basic block, or a file), automated logging must decide
\textbf{where} to place logs, \textbf{what} to record, and \textbf{how} to phrase messages in a readable and maintainable way.
A log statement is a hybrid artifact that combines a code-level logging API call (e.g., \texttt{logger.info(\ldots)}) with a concise natural-language message and selected program variables.
Effective generation, therefore, requires aligning program behavior with project-specific conventions, such as naming, verbosity, and severity-level usage, typically implemented through standard logging frameworks.
This makes automated logging not only a code-generation problem, but also an observability-design problem shaped by developer intent, code context, message readability, and runtime diagnostic value~\cite{fu2014wheredolog,zhu2015learningtolog,Li2023LogMessageReadability,Rong2023LoggingPractices}.

\begin{lstlisting}[style=logcode, language=Java,
caption={A simplified example of single-line logging in Java}, label={lst:logging-example}]
public User getUser(String userId) {
    logger.info("Fetching user with id={}", userId);
    try {
        User user = userRepository.find(userId);
        if (user == null) {
            logger.warn("User not found: id={}", userId);
        }
        return user;
    } catch (DatabaseException ex) {
        logger.error("Failed to fetch user: id={}, error={}", userId, ex.getMessage());
        throw ex;
    }
}
\end{lstlisting}

Listing~\ref{lst:logging-example} illustrates a simplified Java method instrumented with three log statements, and Listing~\ref{lst:logging-output} shows the corresponding runtime logs.
The method logs an \texttt{INFO} message before accessing the repository, emits a \texttt{WARN} when the requested user is missing, and records an \texttt{ERROR} upon a database exception.
Each statement pairs a concise event description with a small set of diagnostic variables (e.g., the user identifier and exception message), enabling developers to reconstruct key control-flow decisions and failure context from the log stream.

\begin{lstlisting}[style=logcode, language={},
caption={Example single-line runtime logs emitted by Listing~\ref{lst:logging-example}},
label={lst:logging-output}]
INFO  2025-05-12 10:15:03 UserService - Fetching user with id=U1024
WARN  2025-05-12 10:15:03 UserService - User not found: id=U1024
ERROR 2025-05-12 10:15:04 UserService - Failed to fetch user: id=U1024, error=connection timeout
\end{lstlisting}

In practice, logs are commonly placed at execution milestones, boundary interactions with external components, and exception-handling paths, where recorded context is most useful for diagnosis~\cite{fu2014wheredolog,zhu2015learningtolog}.
Most systems rely on standard logging frameworks that provide unified APIs and configurable formatting. In Java, widely used options include \texttt{java.util.logging}~\cite{jul}, Log4j~2~\cite{log4j2}, SLF4J~\cite{slf4j}, and Logback~\cite{logback}, while Python typically uses the built-in \texttt{logging} module~\cite{pythonlogging}, sometimes complemented by \texttt{structlog}~\cite{structlog} or \texttt{loguru}~\cite{loguru}.
Accordingly, automated logging techniques generally assume these APIs and aim to generate statements that can be directly inserted and compiled in real projects.

\subsection{Challenges}
Although logging is essential for observability, producing high-quality logs in practice is non-trivial.
Prior automated approaches, typically based on heuristics, static analysis, and traditional machine-learning models, have therefore faced persistent limitations.
Existing studies consistently suggest that logging usefulness is largely determined by three intertwined decisions: where to log, what to log, and how to keep logs effective over time~\cite{fu2014wheredolog,zhu2015learningtolog,proactive2025ase,Zhong2025LogUpdater}.

\subsubsection{Where-to-Log: Placement and Coverage}
A core difficulty of logging is deciding \textbf{where} a statement should be inserted.
Logs placed too sparsely may miss critical runtime evidence, reducing diagnosability, whereas overly dense logging introduces noise, log flooding, and runtime overhead~\cite{fu2014wheredolog,zhu2015learningtolog,autologging2023nier,proactive2025ase}.
In real systems, diagnostically important events often occur along complex, low-frequency control-flow paths (e.g., rare branches, retries, and exception chains), making it difficult to identify insertion points that reliably capture true execution states~\cite{fu2014wheredolog,zhu2015learningtolog,Li2024GoStatic}.

Traditional automated techniques often suggest logging-statement placement using local syntactic cues (e.g., method entry/exit, catch blocks, API calls) or handcrafted rules over AST patterns~\cite{zhu2015learningtolog,mastropaolo2022using,Xie2024FastLog,Xu2024UniLog}.
Such signals are inexpensive and scalable, but they are weak proxies for diagnostic value: they may over-instrument common patterns while missing semantically important events that span multiple statements or methods~\cite{Li2024EffectivenessLLMLogging,Tan2025ALBench}.
Static-analysis-based methods can improve precision by tracking control flow and exceptional paths, yet they still struggle to decide which program points are worth logging from a human-debugging perspective, and they often produce redundant locations when applied at scale~\cite{fu2014wheredolog,Li2024GoStatic,Duan2025PDLogger}.
Moreover, placement decisions made in isolation are difficult to reconcile globally: a log that appears useful within one method may be inconsistent or redundant when viewed across the system, especially in large-scale or distributed applications~\cite{fu2014wheredolog,zhu2015learningtolog,Li2024GoStatic,larger2025arxiv}.

\subsubsection{What-to-Log: Level, Message, and Variables}
Even if a location is chosen, determining \textbf{what} to record remains challenging.
First, log levels encode severity and expected frequency, but these levels are often ambiguous in real projects. The same event may be viewed as \texttt{INFO} in one module and \texttt{WARN} in another, leading to inconsistency and unreliable filtering during incident response~\cite{zhu2015learningtolog,li2021deeplv,Heng2025LogLevelSuggestion}.
Second, messages must convey clear event semantics, yet developers frequently struggle to express intent concisely, avoid vague wording, and align with project terminology~\cite{zhu2015learningtolog,Li2023LogMessageReadability}.
Third, selecting which variables to include is subtle: missing key variables reduces diagnostic value, whereas logging too many values increases verbosity~\cite{zhu2015learningtolog,Li2024GoStatic,Duan2025PDLogger}.
These issues are exacerbated by the need to maintain a consistent style across contributors, components, and versions.

Prior approaches typically decompose this decision into separate subtasks, such as level prediction and variable selection, using local features (identifiers, types, nearby API calls, shallow data-flow signals) and then fill messages with templates~\cite{zhu2015learningtolog,li2021deeplv,Li2024EffectivenessLLMLogging}.
This decomposition is convenient for modeling, but it often yields incoherent outcomes, for example, a predicted level that does not match the message phrasing or variables that are only weakly related to the event being described.
Template- and pattern-based message generation further struggles to express nuanced intent (e.g., distinguishing a transient network hiccup from a persistent configuration error), producing rigid or generic strings that provide limited diagnostic cues.
Overall, the fundamental difficulty is that level, variables, and message are semantically coupled decisions, whereas many traditional techniques treat them as largely independent.

\subsubsection{How-to-Log: Quality, Maintainability, and Evolution}
Logging quality is not only about writing a single statement correctly, but also about keeping logs useful throughout software evolution.
As code and requirements change, logs can become stale, misleading, or incorrect (e.g., referring to outdated conditions, variables, or execution phases), yet they are rarely updated with the same rigor as functional code~\cite{Zhong2025LogUpdater,EvLog2023,Wang2025Defects4Log}.
In addition, practical logging must respect non-functional constraints, such as performance overhead, storage budget, and privacy/security policies, which further complicate the amount and type of information that can be safely recorded~\cite{he2021survey,larger2025arxiv}.

Traditional tooling provides limited support for maintaining logs over time.
Simple consistency checks (e.g., unused variables, formatting rules) can catch surface-level issues, but they do not reliably detect semantic drift, in which the code behavior changes while the log message remains plausible~\cite{Zhong2025LogUpdater,EvLog2023}.
Likewise, mining- or rule-based detectors often rely on project-specific patterns and are brittle in the face of refactoring, API evolution, or style changes~\cite{zhu2015learningtolog,Zhong2025LogUpdater,proactive2025ase}.
As a result, projects accumulate redundant, inconsistent, or outdated logs, gradually reducing the overall utility of the logging infrastructure and increasing maintenance costs~\cite{proactive2025ase,Zhong2025LogUpdater}.

\begin{table*}[t]
  \centering
  \caption{Summary of LLM-based automated logging studies organized by task category.}
  \label{tab:llm_logging_split}
  \resizebox{\textwidth}{!}{
    \begin{tabular}{l l l p{2.0cm} p{2.0cm} p{2.8cm} p{2.2cm} p{2cm} l}
      \toprule
      \textbf{Paper} & \textbf{Task} & \textbf{Paradigm} & \textbf{Input} & \textbf{Output} & \textbf{Context/Technique} & \textbf{Base Model(s)} & \textbf{Comparisons} & \textbf{Metric(s)} \\
      \midrule

      \multicolumn{9}{l}{\textit{\textbf{Full Logging Generation (General Purpose: Location, Message, Variables)}}} \\
      \midrule
      LANCE~\cite{mastropaolo2022using} & Full Gen & Fine-tuning & Java Method & Log Stmt & - & T5 & - & Acc, BLEU \\
      FastLog~\cite{Xie2024FastLog} & Full Gen & Fine-tuning & Method + AST & Insert Point & AST Locator & PLBART & - & Acc, BLEU \\
      LEONID~\cite{mastropaolo2024log} & Multi-Log & Fine-tuning & Full Method & 0-N Logs & - & T5 & - & BLEU, METEOR \\
      ELogger~\cite{Fu2024ELogger} & Block Gen & Fine-tuning & Code Block & Single Log & Block Features & Enc-Dec & - & - \\
      UniLog~\cite{Xu2024UniLog} & Full Gen & ICL & Method & Loc+Level+Msg & Warm-up Demos & Codex, GPT-3 & - & PA, LA, MA \\
      SCLogger~\cite{Li2024GoStatic} & Full Gen & ICL + CoT & Method & Log Stmts & Call Graph (2-hop) & GPT-4, LLaMA2 & GPT-3.5 & PA, AOD, F1 \\
      PDLogger~\cite{Duan2025PDLogger} & Full Gen & ICL + CoT & Method + Slice & Logs & Backward Slicing & o3-mini, DeepSeek & LLaMA3-70B & PA, L-ACC \\
      AUCAD~\cite{Zhang2025AUCAD} & Align Gen & Fine-tuning & Method + Issue & Logs & Issue Alignment & LLaMA3.1 & Magicoder & PA, LA, VF1 \\
      Zhong et al.~\cite{larger2025arxiv} & Full Gen & Fine-tuning & Java Method & Logs & Compact Model FT & LLaMA-8B, Mistral & GPT-4o, Claude & PA, LA, F1 \\

      \midrule
      \multicolumn{9}{l}{\textit{\textbf{Specialized Logging Generation (Test Code, File-level, Configuration)}}} \\
      \midrule
      File-Level~\cite{RuizRodriguez2025FileLevelLogging} & File Gen & ICL & Python File & Multi Logs & Pipeline Stages & GPT-4o-mini & - & L-ACC, AOD \\
      Test-Code~\cite{Shu2025TestCodeLogging} & Test Log & ICL & Test Method & Assert/Log & Test Intent & GPT-3.5, CodeLlama & GPT-4o, Llama3 & Acc, BLEU \\
      ConfLogger~\cite{Shan2026ConfLogger} & Config Log & ICL + Rules & Config Code & Logs & Config Analysis & GPT-4o & - & LA, AOD \\

      \midrule
      \multicolumn{9}{l}{\textit{\textbf{Benchmarks, Empirical Studies, and Visions}}} \\
      \midrule
      LogBench~\cite{Li2024EffectivenessLLMLogging} & Study & ICL / Zero & Method & Level+Msg & - & GPT-3.5/4 & InCoder, StarCoder & BLEU, ROUGE \\
      AL-Bench~\cite{Tan2025ALBench} & Benchmark & - & Java Code & Logs & Benchmark Suite & - & - & ALD, DEA \\
      Auto-Logging~\cite{autologging2023nier} & Vision & - & - & - & AI Instrumentation & - & - & - \\

      \midrule
      \multicolumn{9}{l}{\textit{\textbf{Specific Log Tasks (Level Prediction, Defect Detection, Repair, Quality)}}} \\
      \midrule
      Heng et al.~\cite{Heng2025LogLevelSuggestion} & Level Pred & FT / Zero & Snippet & Level & - & CodeLlama-13B & BERT, RoBERTa & Acc, AUC \\
      OmniLLP~\cite{Ouatiti2025OmniLLP} & Level Pred & RAG + ICL & Log + Neighbors & Level & Cluster Retrieval & CodeXEmbed & - & ARI, AUC \\
      LogUpdater~\cite{Zhong2025LogUpdater} & Log Repair & Agent & Defect Log & Fixed Log & Defect Taxonomy & CodeT5+, GPT-4o & Claude3.5 & BLEU, ROUGE \\
      Defects4Log~\cite{Wang2025Defects4Log} & Detection & ICL + CoT & Method + Log & Defect Type & Defect Patterns & DeepSeek-R1 & GPT-4o & Precision, Recall \\
      LOGIMPROVER~\cite{proactive2025ase} & Quality & - & Codebase & Improved Logs & Proactive Analysis & - & - & - \\

      \bottomrule
    \end{tabular}
  }
\end{table*}

\subsection{LLM-based Approaches for Logging Statement Generation}
\label{subsec:logging-llm}

Table~\ref{tab:llm_logging_split} groups recent LLM-based logging studies by task category, while the discussion below emphasizes the methodological paradigms behind them.
Across these paradigms, the key idea is to exploit LLMs as joint models over code and natural language, so that the three core decisions, where to log, what level to use, and how to phrase the message and variables, are no longer made by hand-crafted rules or local classifiers. These logging decisions emerge from learned semantic patterns and in-context reasoning.
Below, we organize prior work by how it uses LLMs and highlight the mechanisms that make each paradigm effective.

\subsubsection{Fine-tuned End-to-End Generators}
Early work treats logging as a sequence-to-sequence learning problem and fine-tunes encoder--decoder models to transform raw code into logged code.
LANCE~\cite{mastropaolo2022using} fine-tunes a T5 model on pairs of Java methods with and without logging, so that the model learns to rewrite an input method into an output method containing an injected log statement.
During decoding, the model jointly decides where to insert the statement, which level to use, which variables to log, and how to phrase the message.
This joint modeling is precisely where LLMs outperform handcrafted AST patterns: instead of matching rigid templates, the fine-tuned model learns the distribution of logging patterns from large corpora and can adapt to unusual constructs, such as nested exceptions or multi-call sequences, without additional rules.

As the first end-to-end LLM-based automated logging tool, LANCE delivers an innovative proof of concept. Yet, it suffers from three key limitations: (1) rewriting the entire method may inadvertently alter non-logging code, (2) full-method input and output incurs substantial latency compared to local insertion, and (3) it always injects exactly one log statement, thus failing to decide when logging is unnecessary or when multiple logs are required.
Subsequent approaches explicitly refine the use of fine-tuning to mitigate these issues.
FastLog~\cite{Xie2024FastLog} decouples location from content: a learned module first predicts fine-grained insertion points, and the LLM then generates only the log statement to be inserted.
This changes the learning problem from ``rewrite the whole method'' to ``predict a small edit'' and thus preserves the original code while reducing decoding cost.
LEONID~\cite{mastropaolo2024log} extends LANCE by teaching the model to predict not just the content of a single log, but also whether a method needs logging at all and how many logs to insert.
This turns the fixed one-log-per-method assumption into a learned, method-level decision, closer to how developers actually log.
ELogger~\cite{Fu2024ELogger} pushes the same idea to block level: it first classifies whether a block warrants logging and only then invokes the generator, reducing over-instrumentation by using the LLM where it matters most.
AUCAD~\cite{Zhang2025AUCAD} further strengthens this fine-tuning line by automatically constructing aligned datasets from log-related issues. The model is trained not just on arbitrary method/log pairs, but on positive and negative examples mined from real bug reports, which sharpens its notion of “good’’ versus “bad’’ logging behavior.

Overall, these fine-tuned generators demonstrate that LLMs can implicitly encode common logging idioms and jointly model location, level, message, and variables.
Their main trade-off is training and deployment cost: models are tied to specific architectures and datasets, motivating lighter-weight paradigms such as in-context learning.

\subsubsection{Prompt-based In-Context Learning}
With stronger LLMs, an alternative to task-specific fine-tuning is prompt-based in-context learning (ICL).
Compared with fine-tuned models, ICL-based approaches greatly reduce training and infrastructure costs and make it easier to deploy or swap models in practice.
UniLog~\cite{Xu2024UniLog} is the first general-purpose framework to systematically exploit ICL for logging.
It feeds the target method, along with a small set of (code, log) examples from the same project, and asks a Codex-like model~\cite{chen2021codex} to output the location, level, and message.
Mechanistically, UniLog leverages an emergent ability of LLMs: given a handful of exemplars, the model can infer a project’s logging style (e.g., how verbose to be, which variables to mention) and apply that style to new methods.
A warm-up strategy stabilizes this behavior by first exposing the model to generic logging patterns before switching to project-specific examples, reducing variance across invocations.

ICL-based methods shift the burden from training to prompting: instead of adjusting millions of parameters, practitioners adjust the prompt.
This makes automated logging much easier to deploy or migrate between models, and it directly leverages LLM strengths in style transfer and pattern imitation.
File-level ML logging~\cite{RuizRodriguez2025FileLevelLogging} prompts GPT-4o-mini with Python pipeline files and a few annotated examples, letting the model infer where to place logs at pipeline boundaries.
Test-code logging~\cite{Shu2025TestCodeLogging} supplies test methods and exemplar test logs so that the model learns to describe test intent and failure conditions.
ConfLogger~\cite{Shan2026ConfLogger} combines static analysis to find configuration-sensitive operations with ICL prompts that show how such operations are typically logged, enabling the LLM to specialize its suggestions for configuration diagnosability.

\subsubsection{Execution-Aware Reasoning with CoT and Static Context}

While plain ICL uses only the immediate method body, real logging decisions often depend on broader execution context and on a chain of reasoning (e.g., ``this call propagates an error from a lower-level component, so it should be logged at \texttt{ERROR}'').
To expose this context and reasoning to the LLM, several works enrich prompts with chain-of-thought (CoT) style structure and execution-aware static information.

SCLogger~\cite{Li2024GoStatic} augments the input with two-hop callers and callees on the call graph and organizes the prompt in a CoT fashion.
Instead of giving the LLM only the target method, SCLogger explicitly injects the surrounding static context: which functions call this method, what it calls, and how logging is used there.
This design exploits a key LLM capability: when given a broader static context, the model can infer variable roles (e.g., which ID identifies a request versus a user), propagate logging conventions across the call chain, and better align with project-specific patterns without extra training.
The CoT-style prompt further nudges the model to ``think through'' why a location is important before emitting a final log statement.

PDLogger~\cite{Duan2025PDLogger} makes the execution link even tighter by performing backward slicing from predicted log locations.
Data and control dependencies along the slice are turned into explicit prompt tokens, and variables flowing into the logging site are enumerated with their syntactic and semantic roles.
In effect, PDLogger transforms implicit runtime behavior (which values influence this point, under which conditions) into an explicit context for the LLM.
This allows the model to simulate execution more accurately and to generate multiple log statements per method that are consistent with the underlying control and data flow, instead of relying solely on local patterns.

These CoT- and execution-aware prompts thus serve as a bridge between classical program analysis and LLM reasoning: static analyses compute precise dependencies, and the LLM uses them to perform high-level semantic reasoning and message generation.

\subsubsection{Retrieval-Augmented Logging and Subtasks}

Some prior research focuses on specific subtasks (e.g., level prediction, message reconstruction) and combines LLMs with retrieval-augmented generation.

For log levels, Heng et al.~\cite{Heng2025LogLevelSuggestion} benchmark a range of models in zero-shot, few-shot, and fine-tuned settings, showing that local code context suffices in many cases but that borderline severity levels remain difficult.
OmniLLP~\cite{Ouatiti2025OmniLLP} takes a retrieval-augmented view: it clusters log statements by semantic similarity and developer ownership, then retrieves representative neighbors to include in the prompt when predicting the level for a new log.
Here, retrieval plays to LLM strengths in analogical reasoning: instead of deciding ``in the abstract'' whether a statement is \texttt{INFO} or \texttt{WARN}, the model compares the target context against a handful of concrete, project-specific examples and chooses the level that best aligns with those patterns.
This improves level consistency and disambiguation for borderline cases, effectively grounding the LLM in real project practices.

LogBench~\cite{Li2024EffectivenessLLMLogging} investigates the message and level subtasks via a fill-in-the-blank formulation.
It constructs paired datasets of original methods (LogBench-O) and transformed-but-functionally-equivalent methods (LogBench-T), removes existing log statements, and asks LLMs to reconstruct the level and message at fixed locations.
This approach isolates ``what-to-log'' from ``where-to-log'' and assesses whether models rely on brittle syntax or genuine semantic understanding: if performance holds on LogBench-T, where surface syntax has changed, the model is likely relying on deeper semantic cues.
By comparing a variety of commercial and open-source LLMs under this setup, LogBench clarifies which architectures and prompting strategies are better at reconstructing realistic log content.

\subsubsection{Agentic Pipelines for Log Maintenance}

Finally, some work goes beyond generating new logs by using LLM agents in multi-step maintenance pipelines.
LogUpdater~\cite{Zhong2025LogUpdater} first uses a classifier (trained on synthesized defective logs) to detect and categorize logging defects, and then invokes an LLM such as CodeT5+ or GPT-4o to repair the flagged statements.
The approach decomposes the task: a lightweight detector narrows down suspicious logs, and the LLM focuses on high-level semantic repair (e.g., aligning message text with code behavior and fixing mismatched variables or temporal descriptions).
By separating ``where is something wrong?'' from ``how should it be rewritten?'', LogUpdater reduces the risk of unconstrained LLM edits and leverages the model’s strength in natural-language revision instead of raw anomaly detection.

Defects4Log~\cite{Wang2025Defects4Log} complements this pipeline view by providing a benchmark and taxonomy for log defects and systematically evaluating LLMs’ ability to classify and reason about them.
Together, these efforts illustrate a broader pattern: instead of using LLMs as single-shot generators, logging research increasingly wraps them in agentic workflows---detect, classify, explain, repair---so that each step can be checked, constrained, or retried, and logs can evolve along with the codebase.

In summary, LLM-based approaches for logging differ not only in which tasks they tackle, but also in how they expose code, context, and prior examples to the model.
Fine-tuned generators encode logging patterns into model parameters and jointly model location, level, message, and variables, which helps with where-to-log and what-to-log decisions but increases training cost and reduces portability.
ICL methods exploit style imitation and adapt quickly to project conventions, but are sensitive to prompt and example quality.
CoT, static context, and slicing ground placement and variable choices in control/data-flow evidence, improving semantic grounding at the cost of additional system complexity.
Retrieval-augmented subtasks improve consistency for levels and messages through project-specific neighbors, but depend on retrieval quality.
Agentic pipelines support how-to-log and maintenance by decomposing detection, classification, explanation, and repair, but introduce more moving parts and validation needs.
These mechanisms collectively explain why LLMs can address long-standing challenges in where-to-log, what-to-log, and how-to-maintain-logs that were difficult to handle with hand-crafted rules or traditional ML alone.

\rqboxc{Across studies, the key shift introduced by LLMs is not raw generation capability but the ability to treat logging intent as a semantic inference problem instead of a syntactic insertion task. The most effective approaches consistently constrain and inform LLMs using execution-aware context, project-specific exemplars, or multi-step control, instead of relying on unconstrained generation. At the same time, LLMs do not eliminate the ambiguity of logging decisions but relocate difficulty from rule design to context selection, validation, and oversight, pointing toward hybrid, iterative maintenance workflows as a common design pattern.}

\subsection{Performance Evaluation Metrics}
For automated logging, evaluation has gradually shifted from ``exact-match-centric'' reporting to a more nuanced, component-wise view.
Early end-to-end generators often borrowed machine translation-style metrics (e.g., n-gram overlap such as BLEU~\cite{papineni2002bleu}) because the task was framed as producing a single reference-like statement for each context.
However, as LLM-based methods moved to prompt-based generation and multi-log settings, two issues became hard to ignore:
(1) logging is inherently under-specified (multiple insertion points or phrasings can be equally reasonable), and
(2) surface-form metrics are brittle (semantically correct messages can score poorly due to paraphrasing).
As a result, more recent evaluations increasingly combine strict metrics (for comparability) with softer, semantics-aware ones (for faithfulness), such as embedding-based similarity metrics like BERTScore~\cite{zhang2019bertscore}.

Below, we summarize the most widely adopted metrics for each aspect and highlight what they capture (and what they miss).

For \textbf{logging-statement position}, the de facto metric is \textbf{Position Accuracy (PA)}: whether the predicted insertion point matches the reference point at an agreed granularity (often statement/block instead of exact line).
PA is easy to interpret and facilitates comparability across papers, so it remains a useful headline number.
However, PA becomes less informative once methods are allowed to output multiple logs or ``no log'' decisions, because the problem is no longer a single-label prediction.
In such cases, it is more appropriate to treat placement as a set prediction problem and report \textbf{precision/recall/F1} over predicted logging-statement position.
This directly exposes two practically relevant failure modes: over-instrumentation (low precision) and under-instrumentation (low recall).

For \textbf{log level}, \textbf{Level Accuracy (LA)} (or L-ACC) is the simplest and most commonly reported metric.
Yet log levels are not purely categorical in practice: they have an ordinal meaning (e.g., predicting \texttt{WARN} when the reference is \texttt{ERROR} is typically less harmful than predicting \texttt{DEBUG}).
Therefore, modern evaluations often complement LA with an ordinal-distance metric, such as \textbf{Average Ordinal Distance (AOD)}, to quantify the extent to which predictions deviate along a severity scale.
In addition, because log levels can be imbalanced (e.g., \texttt{INFO} dominates), macro-F1 or per-class reporting is often more diagnostic than a single LA number when the goal is to understand behavior on rare but critical levels.

For \textbf{log message}, exact-match metrics such as \textbf{Message Accuracy (MA)} are increasingly treated as a ``strict sanity check'' instead of the main quality signal, since natural language admits paraphrases.
To maintain backward compatibility, many studies still report overlap-based metrics such as \textbf{BLEU/ROUGE}, but these metrics mainly reflect surface similarity and can penalize semantically faithful rephrasing \cite{papineni2002bleu}.
For LLM-generated messages, it is now common (and more defensible) to additionally report a semantics-aware metric such as \textbf{BERTScore}, which compares contextual embedding similarity instead of exact token matches \cite{zhang2019bertscore}.
Finally, edit-distance-style measures (token/character Levenshtein distance) are useful when the paper’s objective includes developer fixing effort: they approximate how much manual editing is needed to correct a generated message.

For \textbf{variables}, the field has largely converged on viewing variable logging as a set selection problem.
Accordingly, \textbf{Variable Precision/Recall/F1} is often more informative than a single ``all-or-nothing'' accuracy.
These metrics separate ``missing important variables'' (low recall) from ``logging too many irrelevant variables'' (low precision), and they also interact naturally with hallucination checks (e.g., penalizing variables that do not exist in scope).
In addition, if a method jointly generates message and variables, it is worth reporting variable metrics separately from message metrics to avoid conflating semantic description quality with instrumentation completeness.
Overall, automated-logging results are not fully comparable across papers because studies differ in dataset construction, insertion granularity, task decomposition (full generation versus subtask prediction), whether multiple valid logs/messages are accepted, programming language and project context, and metric choice.

%% file: Log_Parsing.tex
\section{Log Parsing}
\label{sec:Log_parsing}

Modern software systems generate massive volumes of execution logs to support debugging, monitoring, and reliability engineering.
However, the scale and velocity of log streams make manual inspection impractical: operators must quickly locate relevant events, correlate symptoms across components, and identify anomalous behaviors under strict time constraints.
This motivates automated log analysis, which treats logs as machine-processable data instead of free-form text.
A prerequisite for most downstream log analytics is log parsing, the process of converting raw log lines into structured event records~\cite{he2021survey,beck2025system}.
By normalizing heterogeneous messages into reusable templates and extracting runtime parameters, log parsing enables efficient indexing, aggregation, statistical modeling, and cross-run comparison, and thus directly impacts the effectiveness of tasks such as anomaly detection, failure prediction, and root cause analysis.
Parsing is also where log-specific LLM challenges become especially visible: the parser must decide ambiguous template/value boundaries, preserve stable event identities under drift, cover long-tail templates, and still operate under streaming-scale constraints.

We review LLM-based log parsing via reusable paradigms and the approaches that make them practical at scale, highlighting how they address the limitations of traditional parsers (format drift, long-tail events) and what new trade-offs they introduce (e.g., context budgets and coordination overhead).

\subsection{Task Definition}
\label{subsec:logparsing-task}

Log parsing is the task that transforms each raw log line into a structured representation that separates \textbf{invariant} event semantics from \textbf{dynamic} runtime values.
Concretely, given a raw log line $x$, a parser typically outputs (i) a set of \textbf{header fields} (e.g., timestamp/date, severity level, process/thread, component), (ii) a \textbf{log template} $t$ where dynamic spans are replaced by placeholders (e.g., \texttt{<*>}), and (iii) a \textbf{parameter list} $v$ that records the extracted runtime values.
At the stream level, especially in online settings where logs must be parsed on-the-fly as they arrive, parsers often maintain a \textbf{template inventory} (or event dictionary) and assign an \textbf{event ID} to each unique template so that the raw stream can be converted into an event sequence for downstream mining~\cite{drain2017icws,choi2022logstamp,inferlog2026icse}.

\begin{figure}
  \centering
  \includegraphics[width=0.75\columnwidth]{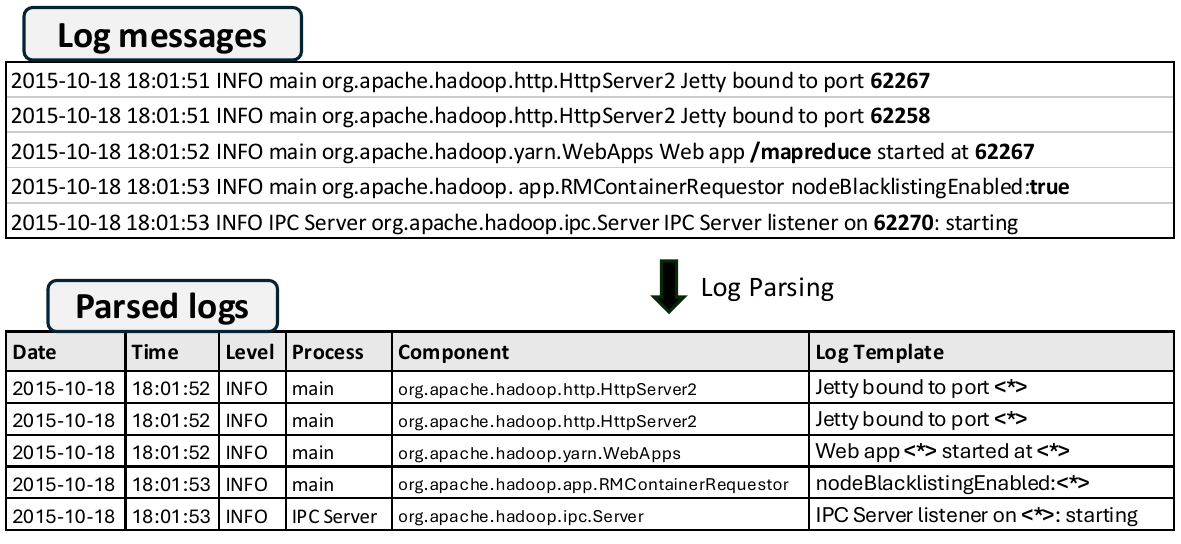}
  \caption{An example log parsing result from Hadoop.}
  \label{fig:parsing}
\end{figure}

Figure~\ref{fig:parsing} illustrates this objective with Hadoop logs.
For example, raw messages such as ``Jetty bound to port 62267'' and ``Jetty bound to port 62258'' differ only in the concrete port number, yet they represent the same underlying event.
A correct parser normalizes them to the same template ``Jetty bound to port \texttt{<*>}'' and extracts the port number as a parameter.
Similarly, the message ``Web app /mapreduce started at 62267'' is parsed into the template ``Web app \texttt{<*>} started at \texttt{<*>}'', separating stable keywords (e.g., ``Web app'', ``started at'') from variable tokens (e.g., the path and port).

In practice, log parsing is rarely a single monolithic step. Most pipelines implement it as a two-stage process.
First, a \textbf{preprocessing} stage extracts standard fields and isolates the message body.
This stage is typically implemented using regular expressions or log format specifications because these header fields are usually emitted in a framework-controlled and relatively stable format (e.g., the timestamp/level/thread/component layout is determined by logging libraries and configuration, and thus changes slowly and predictably).
Second, a \textbf{message parsing} stage focuses on the remaining free-text-like body and performs template--parameter separation.
Unlike headers, the message body is largely developer-written and varies widely across components and versions. It mixes natural-language-like phrases with project-specific tokens and runtime values, and it evolves as code changes.
As a result, a single set of regular expressions is rarely sufficient to robustly normalize message bodies without either over-generalizing templates or exploding them into many variants.
Across these implementations, the core goal remains the same: to produce stable templates that support consistent grouping while accurately extracting the parameters needed for diagnosis.

\subsection{Challenges}
\label{subsec:logparsing-challenges}

Although log parsing has been studied extensively, traditional (non-LLM) parsers face persistent difficulties because logs are semi-structured artifacts produced by evolving software.
These difficulties arise from both the \emph{data} (highly variable message styles and continuous drift) and the \emph{operational requirements} (low latency, high throughput, and stable event identities in online pipelines).
Below, we summarize four recurring challenges that limit the robustness of traditional parsers.

\subsubsection{Structural variability and format drift}
Logs exhibit substantial structural variability across systems, components, and logging libraries, and even within the same system, different modules may adopt different writing styles, delimiters, and token conventions.
Many traditional parsers therefore embed implicit assumptions about message structure (e.g., token positions, separators, fixed vocabularies, or stable prefix patterns) to make parsing tractable~\cite{drain2017icws,dai2020logram}.
Such assumptions can work well for the formats they were designed around, but they degrade when encountering unseen layouts, rare formatting choices, or cross-component inconsistencies.

Crucially, this variability is not static: log formats drift over time as code evolves, dependencies are upgraded, configuration changes alter logging layouts, and developers revise wording for readability or diagnostics~\cite{librelog2025icse,li2024lilac,semanticlog2025tse}.
Under drift, an initially accurate template inventory can become stale, causing template fragmentation (the same event being split into multiple templates) or erroneous merges (different events being grouped together).
Rule-based parsers then require continuous maintenance, while data-driven parsers trained on historical logs can suffer when the distribution of templates or vocabulary shifts.

\subsubsection{Ambiguous template--variable boundaries}
Even with a stable structure, deciding which tokens are \emph{invariant} versus \emph{variable} is inherently ambiguous.
Runtime values have various forms (IDs, UUIDs, paths, IP addresses, error codes), and some values may appear constant within a single deployment but should be treated as variables across runs (e.g., stable ports or hostnames in a fixed environment).
Conversely, some numeric tokens or symbols are semantically meaningful constants (e.g., HTTP status classes or well-known error codes) that should remain in templates.

As a result, common heuristics such as ``all digits are variables'' or delimiter-based splitting can lead to either over-general templates (too many placeholders, losing discriminative semantics) or over-specific templates (template explosion).
Both outcomes reduce downstream utility: overgeneralization blurs distinct events, whereas over-specificity fragments event types, harming aggregation and anomaly detection.

\subsubsection{Long-tail patterns and coverage gaps}
Real log streams often follow a heavy-tailed distribution: a small number of frequent templates dominate the volume. In contrast, rare templates may be crucial for diagnosis (e.g., exception paths and failure-handling branches).
Traditional parsers that rely on frequency cues or stable clustering can underperform on the long tail.
Rare patterns may be misclassified into frequent templates (degrading precision) or split into many singletons (degrading recall and inventory usefulness), resulting in poor coverage precisely where operators most need structured signals.

This long-tail effect also interacts with drift and boundary ambiguity: rare events provide fewer repetitive examples to stabilize grouping, making them more vulnerable to mis-parsing when wording changes or when variable boundaries are subtle.

\subsubsection{Operational constraints and stability requirements}
Operational log parsing is commonly performed at a large scale, where millions of log lines must be processed with low latency, high throughput, and predictable resource usage.
Traditional parsers are attractive in production usages because they provide deterministic execution and low per-line cost, but achieving \emph{both} high throughput and high accuracy becomes difficult as structural variability and drift increase.

Moreover, many downstream pipelines assume that event identities remain stable across time windows (e.g., for alerting, dashboards, and anomaly detectors trained on historical event sequences)~\cite{he2021survey,li2024lilac}.
Template churn caused by drift, boundary ambiguity, or long-tail fragmentation can propagate instability into downstream analytics, degrading both monitoring reliability and cross-run comparability.

\begin{table*}[htbp]
  \centering
  \caption{Summary of LLM-based log parsing approaches. The ``LLM Technique'' column describes the main learning/adaptation paradigm, while ``Reuse / Eff.'' records orthogonal scalability, reuse, or system-level efficiency mechanisms.}
  \label{tab:llm_log_parsing}
  \footnotesize
  \renewcommand{\arraystretch}{0.92}
  \resizebox{\textwidth}{!}{
    \begin{tabular}{p{2cm} l l p{5.9cm} p{1.5cm} p{2.5cm} l}
      \toprule
      \textbf{Paper} &
      \textbf{Setting} &
      \textbf{Superv.} &
      \textbf{LLM Technique} &
      \textbf{Reuse / Eff.} &
      \textbf{Base Model(s)} &
      \textbf{Metric(s)} \\
      \midrule

      \multicolumn{7}{l}{\textit{\textbf{Benchmarks and empirical analyses}}} \\
      \midrule
      Le \& Zhang~\cite{chatgptlogparse2023ase} &
      Online & Both &
      Zero-/few-shot prompting study (capability analysis) &
      -- &
      GPT-3.5 &
      PA, GA, ED \\
      Astekin et al.~\cite{comparative2024esem} &
      Online & Both &
      Few-shot prompting comparison across LLMs &
      -- &
      CodeLlama-7B &
      ED, PA, LCS \\
      LogEval~\cite{Cui2024LogEval} &
      Offline & Sup. &
      Benchmark suite (multiple log-analysis tasks incl. parsing) &
      -- &
      GPT-4 &
      PA, ED \\
      LogBase~\cite{logbase2025issta} &
      Offline & -- &
      Benchmark dataset for semantic log parsing &
      -- &
      -- &
      -- \\
      LLMDPP~\cite{llmdpp2025mobisys} &
      Online & Sup. &
      DPP-based diverse sample selection for few-shot parsing &
      -- &
      Flan-T5-small &
      GA, PA \\
      VISTA~\cite{variable2025fse} &
      Online & Sup. &
      Variable-aware ICL (efficiency/cost-oriented study) &
      Cache &
      GPT-3.5-Turbo &
      PA \\
      \midrule

      \multicolumn{7}{l}{\textit{\textbf{Supervised and log-specific model adaptation}}} \\
      \midrule
      GPT-2C~\cite{le2021gpt2c} &
      Offline & Sup. &
      Fine-tuned GPT-style parser (Q\&A formulation) &
      -- &
      GPT-2 &
      F1 \\
      LogStamp~\cite{choi2022logstamp} &
      Online & Sup. &
      Sequence labeling for template/parameter tagging &
      -- &
      BERT-base &
      RI \\
      LogPPT~\cite{xu2023logppt} &
      Offline & Sup. &
      Prompt-based few-shot token labeling &
      -- &
      RoBERTa &
      PA, GA, ED \\
      LLMParser~\cite{llmparser2024icse} &
      Offline & Sup. &
      Few-shot fine-tuning for template generation &
      -- &
      LLaMA-7B &
      PA, GA \\
      Mehrabi et al.~\cite{abdelwahab2024compactft} &
      Offline & Sup. &
      Compact fine-tuning (effectiveness study) &
      -- &
      Mistral-7B &
      MLA, ED, F1 \\
      OWL~\cite{owl2024iclr} &
      Offline & Sup. &
      Log foundation model training (IT operations) &
      -- &
      OWL &
      RI, F1 \\
      PreLog~\cite{jiang2024prelog} &
      Online & Sup. &
      Log foundation model pretraining (log analytics) &
      -- &
      PreLog-140M &
      GA, OA, ED, uPA \\
      Le et al.~\cite{semanticplm2025icse} &
      Online & Sup. &
      Few-shot tuning of semantic LLMs + adaptive caching &
      Adap. cache &
      RoBERTa (125M) &
      GA, PA, FGA, FTA \\
      LogLM~\cite{liu2025loglm} &
      Offline & Sup. &
      Instruction tuning for log tasks &
      -- &
      LLaMA-2-7B &
      RI, F1 \\
      SuperLog~\cite{adapting2025arxiv} &
      Offline & Sup. &
      Continual pretraining + interpretable domain knowledge &
      -- &
      LLaMA-2-7B &
      RI, F1 \\
      \midrule

      \multicolumn{7}{l}{\textit{\textbf{Prompt-based and adaptive ICL parsing}}} \\
      \midrule
      DivLog~\cite{xu2024divlog} &
      Online & Sup. &
      Adaptive few-shot ICL with prompt enhancement &
      -- &
      GPT-3 (Curie) &
      PA, PTA, RTA \\
      Liu et al.~\cite{Liu2024LogPromptICPC} &
      Online & Unsup. &
      Prompt strategies (fixed demos) for online parsing/analysis &
      -- &
      GPT-3.5-Turbo&
      F1 \\
      Xu et al.~\cite{zeroshot2024icws} &
      Online & Unsup. &
      Zero-shot prompting for template extraction &
      Temp. DB &
      GPT-3.5 &
      PA, GA, ED \\
      LogGenius~\cite{loggenius2024icws} &
      Online & Unsup. &
      Zero-shot prompting + synthetic log generation (assistance) &
      -- &
      GPT-3.5-Turbo &
      PA \\
      Zhou et al.~\cite{math12172758} &
      Online & Unsup. &
      Few-shot prompting (with a BERT-based component in pipeline) &
      -- &
      GPT-3.5 &
      GA, MLA, ED \\
      LLM-TD~\cite{llmtd2025ijis} &
      Offline & Unsup. &
      ICL-based template detection for security event logs &
      Temp. cand. &
      OpenChat &
      FGA \\
      Lemur~\cite{lemur2025iclr} &
      Online & Unsup. &
      Entropy-guided clustering + CoT template merging &
      Cluster &
      GPT-3.5&
      Template accuracy \\
      LogRules~\cite{huang-etal-2025-logrules} &
      Offline & Sup. &
      ICL enhanced with rules (case-/rule-based knowledge injection) &
      Rules &
      GPT-4o-mini + LLaMA-3-8B-Instruct &
      PA, GA, ED, FGA, FTA \\
      \midrule

      \multicolumn{7}{l}{\textit{\textbf{Retrieval/refinement-oriented parsing and system-level efficiency layers}}} \\
      \midrule
      LILAC~\cite{li2024lilac} &
      Online & Sup. &
      ICL + hierarchical demo sampling + cache refinement &
      Adap. cache &
      GPT-3.5 &
      PA, GA, FGA, FTA \\
      LogBatcher~\cite{logbatcher2024ase} &
      Online & Unsup. &
      Demonstration-free parsing via retrieval (no labeled demos) &
      Cache &
      GPT-3.5-Turbo &
      GA, MLA, ED \\
      LogParser-LLM~\cite{logparserllm2024kdd} &
      Online & Unsup. &
      Grouping + multi-step prompting (RAG-style) &
      Grouping &
      GPT-3.5-Turbo&
      GA, PA, FGA, FTA, GGD, PGD \\
      SelfLog~\cite{selflog2024issre} &
      Online & Both &
      Group-wise ICL with self-evolutionary merge tree &
      Tree &
      GPT-3.5 &
      GA, PA, PTA, RTA \\
      HELP~\cite{help2024arxiv} &
      Online & Both &
      Embedding clustering + LLM template generation &
      Cluster &
      text-embedding-3-small + Claude-3.5-Sonnet &
      GA, FGA, PA, FTA \\
      LibreLog~\cite{librelog2025icse} &
      Offline & Unsup. &
      Retrieval-augmented parsing + self-reflection refinement &
      Temp. mem. &
      LLaMA-3-8B-Instruct &
      PA, GA \\
      Parse-LLM~\cite{parselmm2025cikm} &
      Online & Unsup. &
      Agentic decomposition with callable tools (header separation, etc.) &
      Tools &
      GPT-4 &
      GA, FGA, PA, FTA \\
      AdaParser~\cite{adaparser2025icpc} &
      Online & Unsup. &
      Self-generated ICL + self-correction &
      Tree &
      GPT-3.5&
      GA, FGA, PA, FTA \\
      Huang et al.~\cite{nolabel2025fse} &
      Online & Unsup. &
      Unsupervised parser with retrieval (no labeled examples) &
      Temp. DB &
      GPT-3.5 &
      GA, FGA, PA, FTA \\
      EPAS~\cite{epas2025icde} &
      Online & Unsup. &
      Asynchronous scheduling of LLM queries (tail-latency aware) &
      Async &
      Llama-3.1-70B-Instruct &
      GA, FGA, PA, FTA \\
      SemanticLog~\cite{semanticlog2025tse} &
      Online & Unsup. &
      Variable-aware template generation for large-scale parsing &
      Cache &
      LLaMA-2-7B &
      GA, PA, FTA \\
      Duan et al.~\cite{entropy2025isssr} &
      Offline & Unsup. &
      LogBatcher-style grouping + LLM template correction &
      Cache &
      GPT-3.5-Turbo &
      GA, FGA, PA, FTA \\
      InferLog~\cite{inferlog2026icse} &
      Online & Unsup. &
      ICL-oriented prefix KV-cache reuse + auto-tuning &
      KV cache &
      Qwen2.5-14B-Instruct &
      PA, PTA, RTA, GA \\
      \bottomrule
  \end{tabular}}
\end{table*}

\subsection{LLM-based Approaches for Log Parsing}
\label{subsec:parsing-llm}

Table~\ref{tab:llm_log_parsing} shows that LLM-based log parsing has expanded from early capability probes into several reusable methodological families.
Across these families, the core idea is to trade brittle, log-specific heuristics (e.g., delimiter rules and fixed similarity thresholds) for semantic generalization: LLMs can better infer which spans belong to the stable event skeleton versus instance-specific parameters, even when surface forms vary across environments and versions.
We distinguish the main learning/adaptation paradigm from orthogonal reuse and efficiency mechanisms: a parser may be supervised, prompt-based, or retrieval/refinement-oriented, while still using caches, grouping, scheduling, template memories, or tool-based decomposition to reduce cost and stabilize online behavior.

\subsubsection{Benchmarks and empirical analyses as the starting point}
Early work primarily asked whether general-purpose LLMs can parse logs at all and how sensitive they are to prompting choices.
Le \& Zhang~\cite{chatgptlogparse2023ase} provides a representative capability analysis of GPT-3.5 (ChatGPT) under zero-/few-shot prompts, showing that seemingly small prompt-format differences can materially change parsing quality.
As the field matured, benchmark-style efforts (e.g., LogEval~\cite{Cui2024LogEval}) began to standardize evaluation and make it clearer \emph{which aspect} improves (e.g., grouping versus precise template extraction), thereby enabling more comparable method design.

A key takeaway from follow-up empirical analyses is that \emph{example selection} and \emph{variable handling} repeatedly emerge as high-leverage factors that affect both accuracy and cost (e.g., LLMDPP~\cite{llmdpp2025mobisys}, VISTA~\cite{variable2025fse}, and comparative studies~\cite{comparative2024esem}).
In other words, the ``benchmark line'' of work did not merely measure performance. It also identified recurring design knobs that later approaches considered.

\subsubsection{Supervised and log-specific model adaptation: training for structure and controllability}
One family treats log parsing as a supervised learning problem and leverages modern LLMs to learn invariant skeletons and variable boundaries from data, often with greater tolerance for noisy token patterns than purely heuristic rules.
A useful way to view this family is \emph{where supervision is placed}: some methods supervise \emph{token-level decisions} (constant vs.\ variable), while others supervise \emph{template generation} directly.
LogPPT~\cite{xu2023logppt} is a label-efficient supervised parser. It fine-tunes a RoBERTa model on a small labeled dataset to perform few-shot token-level parameter labeling, reducing reliance on handcrafted parsing rules while remaining grounded in curated examples.
Complementarily, LLMParser~\cite{llmparser2024icse} represents lightweight supervised adaptation via few-shot fine-tuning, enabling a compact open-source LLM to directly generate templates with high accuracy when limited labels are available.

In parallel, log-specific pretraining and domain adaptation aim to internalize ``log language'' from large corpora, narrowing the gap between natural text and log tokens.
PreLog~\cite{jiang2024prelog} illustrates this direction by pretraining a log-focused model for log analytics tasks (including parsing), providing a stronger initialization than generic LLMs for downstream log understanding.
Overall, supervised paradigms tend to provide more controllable output formats and predictable behavior (often with lower inference-time cost than per-line prompting), but they trade off flexibility and maintenance: they rely on curated labels or domain corpora, and evolving templates may still require re-training or re-alignment to sustain accuracy under drift.

\subsubsection{Prompt-based and adaptive ICL parsing: structured generation without parameter updates}
A second family avoids parameter updates and instead casts log parsing as a \emph{structured prompting} problem: given a raw message, the model is instructed to emit a normalized template (and optionally the extracted parameters) under a constrained output format.
This direction is attractive for rapid deployment because it can work in a zero-shot or a few-shot manner without collecting training data.
For example, Xu et al.~\cite{zeroshot2024icws} show that even a zero-shot LLM can extract templates online when paired with a lightweight template repository to reuse previously derived results.
Beyond one-shot prompting, fixed-demo prompt strategies have been explored to stabilize parsing behavior in streaming settings (e.g., Liu et al.~\cite{Liu2024LogPromptICPC}), and some pipelines further leverage synthetic or auxiliary LLMs to improve downstream parsing quality (e.g., LogGenius~\cite{loggenius2024icws}).

Within prompting, adaptive ICL refines the paradigm by selecting demonstrations on the fly instead of relying on a static prompt.
DivLog~\cite{xu2024divlog} is representative: it dynamically selects relevant labeled examples for each incoming log, turning parsing into a contextual compare-and-contrast process that helps the model separate invariants from variables for the specific message at hand.
Empirical follow-ups suggest that \emph{diversity} in the selected examples and explicit treatment of variables are recurring levers for robustness and cost control (e.g., LLMDPP~\cite{llmdpp2025mobisys} and VISTA~\cite{variable2025fse}).

More broadly, prompt-based parsing effectively converts ``rules in code'' into ``rules in the prompt'', which improves iteration speed and portability across models, but shifts the main technical challenge to \emph{controllability and normalization}:
the approach must guard against over-generalization, inconsistent placeholders, and format violations, which motivates explicit output constraints, prompt templates tailored to log syntax, and restricted decision spaces such as candidate/template detection prompts (e.g., LLM-TD~\cite{llmtd2025ijis}) or rule-augmented prompting (e.g., LogRules~\cite{huang-etal-2025-logrules}).
In effect, adaptive ICL shifts the core engineering problem from ``design a single good prompt'' to ``design a good \emph{retriever/sampler} for prompts'', which improves coverage but introduces a new dependency on selection quality and context-budget constraints.

\subsubsection{Retrieval/refinement-oriented unsupervised parsing: reducing labels via self-generated context and iterative correction}
A third family targets the no-label regime by replacing curated demonstrations with retrieval and refinement loops.
Instead of supplying ground-truth templates as exemplars, these methods retrieve representative logs (or groups of logs) as contextual anchors and let the LLM infer templates from the retrieved context.
LogBatcher~\cite{logbatcher2024ase} is representative: it performs demonstration-free parsing by retrieving or grouping similar logs and querying the LLM only on representative instances, thereby reducing both labeling and query volume.
LibreLog~\cite{librelog2025icse} pushes this direction further with open-source models: it combines retrieval-augmented parsing with iterative self-reflection and a template memory, using refinement to compensate for missing labels while improving efficiency and privacy.
Related work employs the same approach via self-generated ICL and self-correction mechanisms (e.g., AdaParser~\cite{adaparser2025icpc}) or via unsupervised template repositories that accumulate reusable templates over time~\cite{nolabel2025fse}.

A recurring observation is that ``unsupervised'' here is rarely ``single-shot''.
Effectiveness hinges on the quality of upstream grouping/retrieval: if retrieved logs mix multiple templates, the model may output an averaged, over-general template. If retrieval is too narrow, variables may be frozen as constants.
Consequently, many approaches couple retrieval with explicit \emph{grouping and correction} strategies, such as embedding-driven clustering before template generation (e.g., HELP~\cite{help2024arxiv}), entropy- or structure-guided grouping followed by template merging/refinement (e.g., Lemur~\cite{lemur2025iclr} and entropy-based correction pipelines~\cite{entropy2025isssr}), and memory-based reuse to converge toward stable inventories over long streams~\cite{librelog2025icse,nolabel2025fse}.
In effect, these parsers trade manual labels for approach-level iteration: retrieval/grouping proposes candidate structure, while refinement/memory stabilizes templates across time and workload shifts.

\subsubsection{System-level efficiency layers: grouping, caching, scheduling, and agentic decomposition}
Across the paradigms above, many approaches add system-level layers that make LLM parsing feasible in online or high-volume settings.
These layers are orthogonal to the main learning paradigm: a parser may use prompting, ICL, or retrieval/refinement while also caching templates, grouping similar messages, scheduling LLM calls, or using tools to decompose the workflow.

A common pattern is to explicitly \emph{reuse} past results to amortize LLM queries and stabilize outputs.
LILAC~\cite{li2024lilac} exemplifies this engineering pattern: it combines accuracy-oriented ICL (via carefully selected demonstrations) with an adaptive parsing cache that stores and refines templates over time, reducing repeated calls for recurring patterns while improving consistency under streaming workloads.
Some approaches attach a template repository to zero-shot prompting so previously derived templates can be looked up and reused~\cite{zeroshot2024icws}, while unsupervised parsers can rely on a template database to eliminate labeled demonstrations but still exploit reuse across the stream (e.g., Huang et al.~\cite{nolabel2025fse}).
Other approaches reduce open-ended generation by narrowing the search space, e.g., by prompting the model to identify multiple templates from a batch or from candidate structures (LLM-TD~\cite{llmtd2025ijis}), or by coupling semantic modeling with adaptive caching for robustness under drift~\cite{semanticplm2025icse}.

Grouping-based approaches parse only a small set of representative logs and propagate templates to the rest. LogParser-LLM~\cite{logparserllm2024kdd} is representative of this family, using grouping and multi-step prompting to balance template fidelity against query volume.
Beyond grouping, execution-layer optimizations target tail latency and inference overhead.
EPAS~\cite{epas2025icde} frames online parsing as an asynchronous scheduling problem, while other lines optimize prefix/KV-cache reuse or agentic decomposition (e.g., InferLog~\cite{inferlog2026icse} and Parse-LLM~\cite{parselmm2025cikm}).
The key trade-off is \emph{error amplification}: once an incorrect template is cached, propagated, or promoted into memory, reuse layers can spread it widely.
Hence, practical approaches typically require cache update policies, validation/refinement hooks, and sometimes variable-aware reuse strategies that explicitly track parameter slots to avoid template corruption at scale (e.g., SemanticLog~\cite{semanticlog2025tse}).
A useful takeaway is that these approaches treat LLM reasoning as a high-quality but expensive primitive, and invest in coordination layers to maximize structured output per LLM token.

Overall, the method families address different parts of the challenge space.
Supervised and log-specific model adaptation improves controllability and template--parameter boundary learning when labels or domain corpora are available. Prompt-based and adaptive ICL parsing supports rapid deployment when labeled data are limited and log patterns keep changing, but it depends on prompt and example quality. Retrieval/refinement-oriented parsing reduces manual labeling and helps cover less frequent patterns, but it depends on grouping and retrieval quality. System-level efficiency layers improve online throughput and output stability, but they can amplify cached or propagated errors if validation is weak.

\rqboxc{Early log parsing work focused on feasibility and prompting sensitivity, while more recent work increasingly separates two concerns: the learning/adaptation paradigm used to infer templates, and the reuse or coordination mechanisms used to make LLM parsing practical at scale.
  Importantly, these families of approaches coexist. Supervised adaptation remains attractive when labeled logs are available, prompting dominates rapid deployment, retrieval/refinement reduces labeling needs, and system-level reuse is essential for online or mass scale.
Across them, a unifying design principle is to combine LLM inference with lightweight parsing components, sampling, retrieval, grouping, caching, and validation that amortize LLM cost while preserving template stability under drift.}

\subsection{Performance Evaluation Metrics}
\label{subsec:logparsing-metrics}

Evaluation of log parsing has gradually shifted from \emph{message-level} headline scores toward \emph{template-level} metrics, because production logs are heavy-tailed and drift over time: frequent templates can dominate averages and hide failures on rare but diagnosis-critical events. As a result, recent LLM-based parsers increasingly report classic metrics together with template-level ones to better reflect robustness (e.g., LILAC~\cite{li2024lilac}, Le et al.~\cite{semanticplm2025icse}, LogParser-LLM~\cite{logparserllm2024kdd}).

\phead{Classic and still common: GA and PA (but frequency-biased).}
The most widely used metrics remain \textbf{Grouping Accuracy (GA)} and \textbf{Parsing Accuracy (PA)} because they are simple and comparable across studies~\cite{chatgptlogparse2023ase,Cui2024LogEval,comparative2024esem}.
GA evaluates whether each line is assigned to the correct event group, while PA evaluates whether the per-line template exactly matches the oracle~\cite{chatgptlogparse2023ase,Cui2024LogEval}.
However, both are \emph{message-level} averages: frequent templates dominate, so a parser can score high while failing on the long tail or under drift~\cite{li2024lilac,semanticplm2025icse}.
They can also blur ``grouping is correct'' versus ``placeholder boundaries are correct'', motivating template-level complements~\cite{li2024lilac,huang-etal-2025-logrules}.

\phead{Increasingly adopted for robustness: FGA and FTA.}
To mitigate frequency bias, many recent parsers report \textbf{FGA} (\emph{F1 of Grouping Accuracy}) and \textbf{FTA} (\emph{F1 of Template Accuracy}) as \emph{template-level} complements to GA/PA~\cite{li2024lilac,semanticplm2025icse,logparserllm2024kdd,help2024arxiv}.
FGA scores correctness \emph{per template group} (then aggregates via precision/recall-style F1), making rare-template fragmentation/merging visible. FTA is stricter because it also requires the \emph{template text} (static tokens and placeholders) to match, exposing boundary errors that GA/PA can hide~\cite{li2024lilac,huang-etal-2025-logrules}.
In practice, reporting \{GA,PA\}+\{FGA,FTA\} gives both backward comparability and clearer evidence of template-inventory recovery~\cite{li2024lilac,semanticplm2025icse}.

\phead{Diagnostics and niche variants: ED/LCS, GGD/PGD, PTA/RTA, F1, and RI.}
Several studies add \textbf{Edit Distance (ED)} and/or \textbf{Longest Common Subsequence (LCS)} between predicted and oracle templates to quantify \emph{how far} near-misses are (useful for diagnosing systematic boundary mistakes), but these should be treated as secondary signals instead of standalone correctness measures~\cite{chatgptlogparse2023ase,comparative2024esem,Cui2024LogEval}.
Some papers further introduce \textbf{GGD/PGD} to diagnose \emph{error modes}: whether mistakes stem mainly from over-merging/over-splitting (grouping granularity) or from static$\leftrightarrow$variable boundary mismatches in templates~\cite{logparserllm2024kdd}.
A few prompting-based parsers report \textbf{PTA/RTA} as template-level precision/recall variants that emphasize strict template recovery (often stricter than PA)~\cite{xu2024divlog,selflog2024issre}.
When parsing is formulated as token labeling, \textbf{F1} evaluates whether parameter boundaries are identified correctly~\cite{le2021gpt2c}. And for clustering-style evaluation, \textbf{RI} (Rand Index) measures pairwise agreement of the predicted vs.\ oracle partition (together vs.\ apart)~\cite{choi2022logstamp,owl2024iclr}.
Even within log parsing, direct cross-paper comparison remains limited because studies differ in datasets, online versus offline settings, oracle template definitions, grouping policies, and which metrics they report.

%% file: Log_Mining.tex
\section{Downstream Log Analysis Tasks}
\label{sec:Log_analysis}

\begin{figure}[h!]
  \centering
  \includegraphics[width=1\columnwidth]{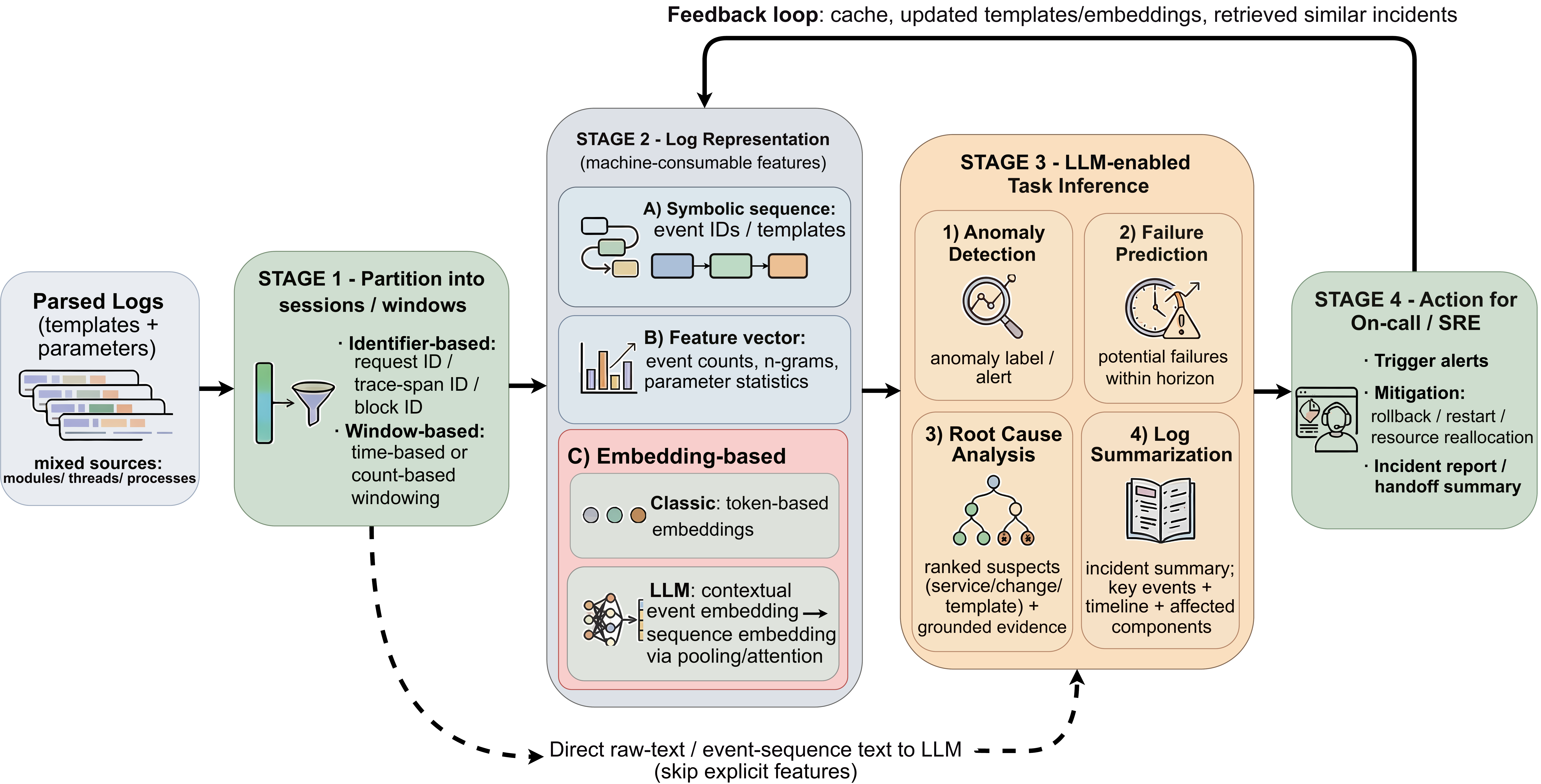}
  \caption{A common workflow for LLM-enabled downstream log analysis.}
  \label{fig:workflow}
\end{figure}

Downstream log analysis aims to turn high-volume log evidence into actionable signals.
Many pipelines operate on parsed templates and parameters, while some LLM-based approaches work directly with raw, lightly normalized, or embedded log windows.
In this survey, we focus on four representative tasks: \textbf{anomaly detection} (flagging abnormal log sequences or windows), \textbf{failure prediction} (forecasting failures within a future time horizon), \textbf{root cause analysis} (identifying the most likely cause or component of an incident and justifying it with evidence), and \textbf{log summarization} (compressing large log windows into concise, operator-friendly summaries).

\phead{Challenges motivating the shift toward LLMs.}
Historically, operators often relied on keyword search and manual inspection, which is fragile under message vocabulary variation and scale~\cite{he2021survey}.
Rule-based analytics can be efficient but are brittle under format drift and hard to adapt across systems, especially when log formats and templates evolve over time~\cite{drain2017icws,dai2020logram}.
Learning-based detectors and predictors improve automation, yet they often depend on careful feature engineering, stable parsing/representation choices, and sufficient labeled data, and can still be sensitive to log evolution~\cite{Ott2021RobustTransferable}.
They may also struggle to provide human-friendly explanations, motivating more interpretable and interaction-oriented log analysis workflows~\cite{Liu2024LogPromptICPC,Karlsen2024BenchmarkingLLMsJNSM}.
In the LLM era, researchers increasingly leverage models’ semantic generalization to (i) better tolerate heterogeneous phrasing and evolution, and (ii) generate explanation-oriented outputs that help operators understand not only \textit{what} happened but also \textit{why}~\cite{Cui2024LogEval,owl2024iclr}.

\phead{A common workflow: partition $\rightarrow$ representation $\rightarrow$ task inference $\rightarrow$ action.}
In practice, downstream analysis is rarely performed on a monolithic log file.
Instead, logs are first \textit{partitioned} into sessions or windows to separate mixed sources (modules/threads/processes) into coherent log sequences.
Partitioning can be driven by identifiers (e.g., request ID, trace/span ID, block ID) or by time- or count-based windowing when identifiers are unavailable.
After sequences are formed, a pipeline may parse events, derive embeddings or other \textit{log representations}, select salient context, or directly prompt an LLM, and then apply task-specific approaches to trigger alerts, rank suspected causes, predict upcoming failures, or summarize incidents for operators.
Figure~\ref{fig:workflow} summarizes this workflow and highlights where LLMs are commonly integrated to improve failure perception, diagnosis, and operator-facing actions.

We organize the rest of this section following the workflow in Figure~\ref{fig:workflow}.
Section~\ref{subsec:logrepr} discusses \textbf{log representation}, which largely determines what any downstream model can learn or infer.
Sections~\ref{subsec:logad}, \ref{subsec:logfp}, \ref{subsec:logrca}, and \ref{subsec:logsum} then cover four representative LLM-enabled tasks:
\textbf{anomaly detection}, \textbf{failure prediction}, \textbf{root cause analysis}, and \textbf{log summarization}.

\subsection{Log Representation}
\label{subsec:logrepr}

Given a sessionized (or windowed) log sequence, typically a sequence of parsed events/templates with optional parameters and timestamps, \textbf{log representation} maps it into machine-consumable features that preserve task-relevant signals while suppressing nuisance variability.
Depending on the downstream task, the representation can be (i) a discrete symbolic sequence (event IDs/templates), (ii) a fixed-length feature vector (counts, n-grams, parameter statistics), or (iii) a continuous embedding sequence or sequence embedding produced by neural encoders.
Logs contribute information that is difficult to replace with metrics, traces, or code alone: symptom wording, exception/error text, component-local execution narratives, chronological evidence trails, and sparse but informative runtime context.
Representation choice is also a major source of cross-paper incomparability because studies differ in whether they use raw messages, parsed templates, event IDs, parameters, embeddings, or direct prompt inputs.
Figure~\ref{fig:logrepr_overview} illustrates a log window and how it can be mapped into different representation types.

\begin{figure}[t]
  \centering
  \resizebox{\linewidth}{!}{
    \begin{tikzpicture}[
        font=\footnotesize,
        box/.style={draw,rounded corners,align=left,inner sep=4pt,minimum height=10mm},
        arr/.style={-{Stealth[length=2mm]},thick}
      ]
      \node[box] (raw) {\textbf{Raw logs (window)}\\
        \texttt{10:12:03.120 INFO  OrderSvc  recv request id=42}\\
        \texttt{10:12:03.128 INFO  OrderSvc  GET /api/orders?id=42}\\
        \texttt{10:12:03.311 WARN  DB        timeout conn=7 (3000ms)}\\
        \texttt{10:12:03.540 INFO  OrderSvc  retry query attempt=1}\\
      \texttt{10:12:03.905 ERROR Gateway   upstream failed code=504}};

      \node[box, below=3mm of raw] (parsed) {\textbf{Parsed events} (templates + parameters)\\
        $S=\langle (e_i,t_i,p_i)\rangle$\\[1mm]
        $e_1$: \texttt{recv request id=*} \quad $p_1{=}\{\texttt{id}=42\}$\\
        $e_2$: \texttt{GET /api/orders?id=*} \quad $p_2{=}\{\texttt{id}=42\}$\\
        $e_3$: \texttt{timeout conn=* (\#ms)} \quad $p_3{=}\{\texttt{conn}=7,\,\texttt{ms}=3000\}$\\
        $e_4$: \texttt{retry query attempt=*} \quad $p_4{=}\{\texttt{attempt}=1\}$\\
      $e_5$: \texttt{upstream failed code=*} \quad $p_5{=}\{\texttt{code}=504\}$};

      \draw[arr] (raw.south) -- node[right]{\textit{parsing}} (parsed.north);

      \node[draw,rounded corners,inner sep=3pt,fit=(raw)(parsed),label=above:{\textbf{Log window}}] (inputbox) {};

      \coordinate (merge) at ($ (inputbox.east)$);

      \node[box, right=10mm of merge, minimum width=35mm] (lane1) {\textbf{Word embeddings}\\
        token vectors $\rightarrow$\\
      avg / TF-IDF};
      \node[box, above=9mm of lane1, minimum width=35mm] (lane2) {\textbf{Classic counts}\\
        bag-of-events / n-grams\\
      param statistics};
      \node[box, below=9mm of lane1, minimum width=35mm] (lane3) {\textbf{Contextual encoder}\\
        BERT/SBERT/\emph{log enc.}\\
      event embedding};
      \node[box, below=9mm of lane3, minimum width=35mm] (lane4) {\textbf{Direct LLM}\\
        raw window text\\
      prompt \& reason};

      \draw[arr] (merge) -- (lane2.west);
      \draw[arr] (merge) -- (lane1.west);
      \draw[arr] (merge) -- (lane3.west);
      \draw[arr] (merge) -- (lane4.west);

      \node[box, right=9mm of lane1, minimum width=44mm] (out1) {\textbf{Word-avg event embs} $\mathbf{z}^{(w)}_i$\\
        $\mathbf{z}^{(w)}_3=\mathrm{avg}(\texttt{timeout},\texttt{conn},\texttt{ms})$\\
        $\approx [0.12,\,-0.03,\,0.44]$\\
      (context-free tokens)};

      \node[box, right=9mm of lane2, minimum width=44mm] (out2) {\textbf{Vector features} $\mathbf{x}$\\
        counts over templates\\
        $[c(e_1),\ldots,c(e_5)] = [1,1,1,1,1]$\\
      + param stats (e.g., $\mu(\texttt{ms})=3000$)};

      \node[box, right=9mm of lane3, minimum width=44mm] (out3) {\textbf{Contextual event embs} $\mathbf{z}^{(c)}_i$\\
        embed full template+params\\
        $\mathbf{z}^{(c)}_3\approx[0.08,\,0.51,\,-0.11]$\\
      (context-sensitive)};

      \node[box, right=9mm of lane4, minimum width=44mm] (out4) {\textbf{Prompt output}\\
        \textit{Likely DB timeout caused 504.}\\
      \textit{Retries observed (attempt=1).}};

      \draw[arr] (lane1.east) -- (out1.west);
      \draw[arr] (lane2.east) -- (out2.west);
      \draw[arr] (lane3.east) -- (out3.west);
      \draw[arr] (lane4.east) -- (out4.west);

      \node[box, right=10mm of out1, minimum width=34mm] (tasks) {\textbf{Downstream tasks}\\
      AD / FP / RCA / Sum};

      \draw[arr] (out1.east) -- (tasks.west);
      \draw[arr] (out2.east) -- (tasks.west);
      \draw[arr] (out3.east) -- (tasks.west);
      \draw[arr] (out4.east) -- (tasks.west);

    \end{tikzpicture}
  }
  \caption{An illustrative overview of log window representations: raw logs are parsed into templates and parameters, then mapped to (i) classic feature vectors, (ii) context-free word-averaged embeddings, (iii) contextual event embeddings, or (iv) direct prompt-based LLM outputs. Numeric vectors are schematic to highlight different representations.}
  \label{fig:logrepr_overview}
\end{figure}

\phead{Traditional representations: statistical, parameter-aware, and semantic sequence features.}
A common starting point is \textbf{statistics-based} representation, which converts each log sequence into a feature vector by counting event occurrences (bag-of-events) or short patterns such as n-grams.
This view is compact and efficient, but it can lose semantic similarity between differently worded events and captures ordering only shallowly.
A closely related line is \textbf{parameter-aware} representation, which augments event statistics with counts/patterns over extracted parameter values (IDs, paths, codes), useful when diagnosis depends on \textit{which entities/values} appear, but potentially brittle under drift and high-cardinality parameters.

The most widely used \textbf{semantic} representations in traditional pipelines build embeddings from the message/template text.
A typical workflow is to map each token in a log event to a word embedding (e.g., word2vec algorithm~\cite{mikolov2013distributed}), aggregate token vectors into an \textit{event embedding} (often by weighted averaging), and then compose a \textit{sequence embedding} by pooling or concatenating embeddings across events in the window.
This family is attractive because it provides a continuous semantic space while remaining lightweight.

\phead{LLM-era representations: reuse classic features or switch to contextual encoders.}
Many recent LLM-based log analytics approaches still reuse classic representations (event-count vectors, parameter statistics, n-grams, or word-embedding-based features) as inputs to downstream models because they are cheap, stable, and compatible with existing monitoring pipelines.
In these hybrid designs, LLMs may serve only as an auxiliary component (e.g., to improve parsing quality or provide explanations), while the anomaly detector/predictor still relies on traditional features and machine-learning models.

At the same time, many recent pipelines adopt \textit{contextual event embeddings}. Instead of averaging context-free word vectors, they embed each entire log event with a pretrained encoder.
The main advantage is robustness: contextual encoders model word order and compositional meaning within an event, and they can assign different representations to the same token across different contexts (which is common in fast-evolving software logs)~\cite{peters-etal-2018-deep,ethayarajh2019contextual}.
As a result, semantically equivalent events with heterogeneous phrasing tend to stay closer in the embedding space, while subtle boundary or wording changes are less likely to break downstream models, a property that is particularly useful under drift, long-tail patterns, and limited supervision.

In practice, the workflow is lightweight.
Each parsed event (often the template text, optionally concatenated with selected parameters) is converted to a short input string and encoded with a contextual model such as BERT~\cite{devlin2019bert} and Sentence-BERT~\cite{reimers2019sentencebert}, or with log-oriented encoders such as PreLog~\cite{jiang2024prelog} and OWL~\cite{owl2024iclr}, to obtain a fixed-length \textit{event embedding}.
For sequence-level tasks, event embeddings within a session/window are then aggregated (e.g., pooling/attention or a lightweight sequential model) into a \textit{window representation}.
Downstream modules use these representations either for (i) similarity-based retrieval/matching (nearest-neighbor search for related events/windows) or (ii) training detectors/predictors for anomaly detection and failure prediction, often with improved tolerance to wording variation compared to word-embedding pipelines.

\phead{Unified log representations via large-scale pretraining.}
Beyond ``using a general-purpose encoder,'' BigLog~\cite{10188759} advocates \emph{unsupervised large-scale pre-training} to learn a \emph{unified} representation space for logs.
The core idea is to pretrain on massive, heterogeneous log corpora so that a single encoder can produce transferable embeddings across systems and log styles, reducing the reliance on per-system feature engineering or task-specific representation tuning.
This perspective is especially relevant in LLM-era pipelines where the bottleneck often shifts from model capacity to \emph{representation compatibility}: when logs change frequently or when operators need cross-service retrieval and correlation, a unified embedding space enables (i) more reliable similarity search over historical windows, (ii) easier cross-system transfer/bootstrapping under limited labels, and (iii) a common interface that multiple downstream tasks (anomaly detection, failure diagnosis, summarization) can share.
In practice, BigLog-style pretraining can be viewed as a middle ground between ``classic features'' and ``prompt on raw text'': it keeps inference efficient (fixed-length embeddings) while improving semantic robustness and portability at scale.

\phead{Skipping explicit features: direct raw-text or event-sequence text as input.}
Finally, some LLM-based approaches partially or fully bypass handcrafted feature construction by feeding an LLM the \textit{raw log window} (or a text form of the \textit{event sequence}) and prompting it to perform the downstream task directly.
This design can preserve richer context and enable explanation-oriented outputs, but it introduces new constraints: long contexts require careful windowing, summarization, or retrieval to control noise and cost, and online deployment may demand additional compression or caching mechanisms.
Overall, representation choices trade off efficiency, robustness, and fidelity. Thus, many recent approaches combine compact classic features with contextual embeddings or selective raw-text reasoning, depending on task requirements and operational budgets.

\subsection{Anomaly Detection}
\label{subsec:logad}
\subsubsection{Task definition}
Log anomaly detection identifies \textit{unexpected} or \textit{undesired} system behavior patterns from logs.
Such anomalies typically indicate potential faults, misconfigurations, performance regressions, or other conditions that violate the expected normal execution, and thus serve as early signals for reliability triage.

For anomaly detection, results support log-specific findings in three situations: runtime logs are the main evaluated input, runtime logs are an explicit part of multimodal evidence, or the reported results explicitly involve logs in a setting where log use is optional.

Given a sessionized (or windowed) log sequence
$S = \langle (e_i, t_i, p_i) \rangle_{i=1}^{n}$,
where $e_i$ is a parsed event/template (or the raw log message), $t_i$ is the timestamp, and $p_i$ are optional parameters,
the task outputs either (i) an anomaly score $a(S)$ or a binary label $y \in \{0,1\}$ indicating whether the sequence is abnormal,
and optionally (ii) localized evidence such as suspicious positions $\mathcal{I} \subseteq \{1,\ldots,n\}$ (or key templates/snippets) to support debugging.

In practice, anomaly detection is usually \textit{reactive}: it flags abnormal states or behavior patterns in observed logs (often in unsupervised or weakly supervised settings), without necessarily predicting whether/when a concrete failure will occur.

\begin{table*}[t]
  \centering
  \caption{Summary of LLM-based anomaly detection studies organized by task category. (continued in Table~\ref{tab:ad2})}
  \label{tab:ad}
  \resizebox{\textwidth}{!}{
    \begin{tabular}{l l l p{3.5cm} l l p{2.8cm}}
      \toprule
      \textbf{Paper} & \textbf{Supervised?} & \textbf{Train?} & \textbf{Dataset(s) evaluated} & \textbf{Parsing?} & \textbf{Metric(s)} & \textbf{Base Model(s)} \\
      \midrule

      \multicolumn{7}{l}{\textit{\textbf{Discriminative \& Representation Learning (Encoder-based)}}} \\
      \midrule
      HitAnomaly~\cite{Huang2020HitAnomaly} & Supervised & Yes & HDFS, BGL, OpenStack & Yes & F1 & BERT \\
      SwissLog~\cite{Li2020SwissLog} & Supervised & Yes & HDFS, BGL & Yes & F1 & BERT \\
      Ott et al.~\cite{Ott2021RobustTransferable} & Semi-Sup & Yes & OpenStack & Yes & F1 & BERT \\
      LogBert~\cite{Guo2021LogBERT} & Semi-Sup & Yes & HDFS, BGL, Thunderbird & Yes & F1 & BERT \\
      NeuralLog~\cite{Le2021NoParsing} & Supervised & Yes & HDFS, BGL, TB, Spirit & No & F1 & BERT \\
      BERT-Log~\cite{Chen2022BERTLog} & Supervised & Yes & HDFS, BGL & Yes & F1 & BERT-110M \\
      LogST~\cite{Zhang2022LogST} & Semi-Sup & Yes & HDFS & Yes & F1 & Sentence-BERT \\
      Prog-BERT-LSTM~\cite{shao2022log} & Semi-Sup & Yes & HDFS, BGL, Thunderbird & Yes & F1 & BERT \\
      HilBERT~\cite{Huang2023HilBERT} & Supervised & Yes & HDFS, BGL & Yes & F1 & BERT \\
      LogBP-LORA~\cite{he2023parameter} & Supervised & Yes & BGL, Thunderbird & Yes & F1 & BERT-base \\
      LogDAPT~\cite{Zheng2023LogDAPT} & Supervised & Yes & HDFS, BGL, Thunderbird & Yes & F1 & BERT \\
      LogADSBERT~\cite{Hu2023SentenceBERTLogAD} & Semi-Sup & Yes & HDFS, Hadoop, OpenStack & Yes & F1 & Sentence-BERT \\
      Karlsen et al.~\cite{Karlsen2023SemanticVsSyntactic} & Unsup & No & ECML, CSIC, Access & Yes & F1 & DistilRoBERTa \\
      FastLogAD~\cite{Lin2024FastLogAD} & Semi-Sup & Yes & HDFS, BGL, Thunderbird & Yes & F1 & ELECTRA \\
      LAnoBERT~\cite{Lee2021LAnoBERT} & Semi-Sup & Yes & HDFS, BGL, Thunderbird & No & F1, AUC & BERT \\
      Corbelle et al.~\cite{2024Semantic} & Supervised & Yes & HDFS & Yes & F1 & BERT-base \\
      Zhou et al.~\cite{math12172758} & Supervised & Yes & HDFS, BGL & Yes & F1 & BERT \\
      LogELECTRA~\cite{Yamanaka2024LogELECTRA} & Semi-Sup & Yes & BGL, Thunderbird, Spirit & No & F1 & ELECTRA \\
      LogFiT~\cite{almodovar2024logfit} & Semi-Sup & Yes & HDFS, BGL, Thunderbird & No & F1 & BERT \\
      LogRoBERTa~\cite{sun2026improving} & Supervised & Yes & HDFS, BGL & No & F1 & RoBERTa \\
      ADALog~\cite{Pospieszny2025ADALog} & Unsup & Yes & BGL, Thunderbird, Spirit & No & F1 & DistilBERT \\
      Thali et al.~\cite{thali2025icsllm} & Supervised & Yes & SWaT and Morris ICS telemetry & No & F1 & BERT \\
      APT-LLM~\cite{benabderrahmane2025aptllm} & Semi-Sup & Yes & DARPA TC host-audit events & Yes & AUC & BERT \\
      Entezami et al.~\cite{entezami2025networkanomalies} & Supervised & Yes & KDD99, UNSW, CICIDS network traffic & No & F1 & BERT \\
      AnomalyExplainerBot~\cite{balasubramanian2025anomalyexplainerbot} & Supervised & Yes & HDFS & Yes & F1 & RoBERTa \\
      Horv{\'a}th et al.~\cite{horvath2025realtimelogscale} & Supervised & Yes & BGL, Thunderbird & Yes & F1, Acc & LLaMA \\

      \midrule
      \multicolumn{7}{l}{\textit{\textbf{Generative Forecasting \& Next-Token Prediction (Decoder-based)}}} \\
      \midrule
      LogGPT (BigData)~\cite{Trabelsi2023LogGPT} & Supervised & Yes & HDFS, BGL, Thunderbird & No & F1 & GPT-2 \\
      LogSense~\cite{srivatsalogsense} & Semi-Sup & Yes & - & Yes & - & GPT-2 \\
      PreLog~\cite{jiang2024prelog} & Supervised & Yes & HDFS, BGL, Spirit & No & F1 & FLAN-T5 (140M) \\
      Bogdan et al.~\cite{bogdan2025goodenough} & Supervised & Yes & ECU network packets & No & Top-k & Qwen2-0.4B \\
      LogLM~\cite{liu2025loglm} & Supervised & Yes & BGL, Spirit & Yes & F1 & LLaMA-2-7B \\
      NLPLog~\cite{adapting2025arxiv} & Supervised & Yes & NLPLog & Yes & F1 & LLaMA-2-7B \\
      \bottomrule
    \end{tabular}
  }
\end{table*}

\begin{table*}[p]
  \centering
  \caption{Summary of LLM-based anomaly detection studies organized by task category. (continued from Table~\ref{tab:ad}.)}
  \label{tab:ad2}
  \renewcommand{\arraystretch}{0.85}
  \resizebox{\textwidth}{!}{
    \begin{tabular}{l l l p{3.5cm} l l p{2.8cm}}
      \toprule
      \textbf{Paper} & \textbf{Supervised?} & \textbf{Train?} & \textbf{Dataset(s) evaluated} & \textbf{Parsing?} & \textbf{Metric(s)} & \textbf{Base Model(s)} \\
      \midrule
      \multicolumn{7}{l}{\textit{\textbf{Prompt-Based \& In-Context Learning (Zero/Few-shot)}}} \\
      \midrule
      LogPrompt (IJCNN)~\cite{2023LogPrompt} & Supervised & Yes & HDFS, BGL & No & F1, Acc & BERT-large \\
      LogPrompt (ICPC)~\cite{Liu2024LogPromptICPC} & Unsup & No & HDFS, BGL & Yes & F1 & ChatGPT \\
      Setu et al.~\cite{mannam2023optimizing} & Supervised & Yes & Apache, BGL, HDFS & Yes & F1, AUC & GPT-3 \\
      LogGPT (HPCC)~\cite{Qi2023LogGPTExploringChatGPT} & Supervised & No & BGL, Spirit & Yes & F1 & GPT-3.5-turbo \\
      Egersdoerfer et al.~\cite{Egersdoerfer2023EarlyExplorationChatGPT} & Supervised & No & Lustre & No & F1 & GPT-3.5-turbo \\
      Owl~\cite{owl2024iclr} & New Model & Yes & BGL, Spirit & Yes & F1 & Owl (LLaMA) \\
      Fariha et al.~\cite{fariha2024log} & Unsup & Yes & HDFS & Yes & F1 & GPT-3.5-turbo \\
      VMFT-LAD~\cite{senevirathne2024virtual} & Semi-Sup & Yes & VM/Server failures & Yes & AUC & GPT-3.5-turbo \\
      FLEXLOG~\cite{hadadi2025llm_meets_ml} & Supervised & Yes & ADFA, SynHDFS & Yes & F1 & Mistral-22B \\
      CloudAnoAgent~\cite{zou2025cloudanobench} & Supervised & No & CloudAnoBench (logs + metrics) & No & Acc & Gemini-2.5 \\
      \midrule
      \multicolumn{7}{l}{\textit{\textbf{Retrieval-Augmented Generation (RAG)}}} \\
      \midrule
      RAPID~\cite{No2024RAPID} & Semi-Sup & No & HDFS, BGL, Thunderbird & No & F1 & BERT \\
      LogRag~\cite{2024lograg} & Semi-Sup & Yes & BGL, Spirit & Yes & Prec, Rec, F1 & GPT-3.5-turbo, Mistral-7B-Instruct \\
      RAGLog~\cite{Pan2023RAGLog} & Semi-Sup & No & BGL, Thunderbird & No & F1 & RAG Model \\
      XRAGLog~\cite{zhang2025xraglog} & Semi-Sup & No & HDFS, BGL, Thunderbird & Yes & F1 & GPT-3.5 \\

      \midrule
      \multicolumn{7}{l}{\textit{\textbf{Hybrid \& Collaborative (Small Model + Large Model)}}} \\
      \midrule
      LLMeLog~\cite{he2024llmelog} & Supervised & Yes & HDFS, BGL, Thunderbird & Yes & F1 & ChatGPT + BERT \\
      LogFormer~\cite{Guo2024LogFormer} & Supervised & Yes & HDFS, BGL, TB, GAIA & No & F1 & S-BERT + ChatGPT \\
      CLogLLM~\cite{ren2024clogllm} & Unsup & Yes & HDFS, BGL, Thunderbird & No & F1 & Qwen + GPT-3.5 \\
      LogLLM~\cite{Guan2024LogLLM} & Supervised & Yes & HDFS, BGL, TB, Liberty & No & F1 & BERT + LLaMA-3 \\
      AdaptiveLog~\cite{ma2025adaptivelog} & Supervised & Yes & BGL, Thunderbird & No & F1 & BERT + ChatGPT \\
      LogRules~\cite{huang-etal-2025-logrules} & Supervised & Yes & BGL, Spirit & No & F1 & GPT-4o + LLaMA-3 \\
      LogADRef~\cite{lim2025parameter_efficient_lad} & Supervised & Yes & BGL, HDFS, Spirit, TB & Yes & F1 & RoBERTa + GPT2 \\
      SemiRALD~\cite{sun2025semirald} & Supervised & Yes & HDFS, BGL & Yes & F1 & ChatGPT + BERT \\
      LogSynergy~\cite{sui2025BridgingtheGap} & Supervised & Yes & Three production datasets, BGL, Thunderbird, Spirit & Yes & F1 & ChatGPT-4o \\
      CLSLog~\cite{xiao2025clslog} & Supervised & Yes & BGL, Zookeeper & No & F1 & BERT + Qwen \\
      LEMAD~\cite{ji2025lemad} & Semi-Sup & Yes & State Grid logs + runtime metrics & Yes & F1 & BERT + GPT-4o \\
      CoLA~\cite{zhu2025cola} & Supervised & Yes & HDFS, BGL, Spirit & Yes & F1 & LLaMA-3 + GPT-4o \\
      LUK~\cite{ma2024luk} & Supervised & Yes & BGL, Spirit, Thunderbird & Yes & F1 & GPT-3.5 + Expert \\
      SemiSMAC~\cite{sun2025semismac} & Supervised & Yes & BGL, HDFS, Spirit & Yes & F1 & GPT-4-turbo \\

      \midrule
      \multicolumn{7}{l}{\textit{\textbf{Agentic, Reasoning \& Neuro-Symbolic Frameworks}}} \\
      \midrule
      Audit-LLM~\cite{Song2024AuditLLM} & Unsup & No & CERT r4.2/5.2 & No & F1 & GPT-3.5 (Agent) \\
      WebNorm~\cite{liao2024detecting} & Semi-Sup & No & TrainTicket, NiceFish & Yes & F1 & GPT-3.5 (Agent) \\
      LogRESP-Agent~\cite{2025LogRESP-Agent} & Semi-Sup & Yes & EVTX-ATTACK & No & F1, TPR & Gemini-2.0-Flash \\
      Song et al.~\cite{song-etal-2025-confront} & Supervised & Yes & CERT v6.2 & No & F1 & ChatGLM-6B \\
      SHIELD~\cite{sun2025alerts2intelligence} & Supervised & Yes & DARPA-E3, ATLASv2 & Yes & Prec & DeepSeek-R1 \\
      LogReasoner~\cite{ma2025logreasoner} & Supervised & Yes & BGL, Spirit & No & F1 & Qwen2.5-1.5B \\

      \midrule
      \multicolumn{7}{l}{\textit{\textbf{Data Generation, Benchmarks \& Visions}}} \\
      \midrule
      Karlsen et al.~\cite{Karlsen2024BenchmarkingLLMsJNSM} & Supervised & Yes & CSIC, Thunderbird, BGL & Yes & F1 & RoBERTa/GPT \\
      LogEval~\cite{Cui2024LogEval} & Supervised & No & BGL, Thunderbird & No & F1 & GPT-4 \\
      AD-LLM~\cite{yang-etal-2025-ad} & Semi-Sup & Yes & News, reviews, and spam & No & AUC & LLaMA-3/GPT-4o \\
      AnomalyGen~\cite{li2025anomalygen} & Supervised & Yes & Hadoop, HDFS & Yes & F1 & GPT-4o \\

      \bottomrule
    \end{tabular}
  }
\end{table*}

\subsubsection{LLM-based Approaches for Anomaly Detection}
\label{subsec:logad_llm}

Table~\ref{tab:ad}--\ref{tab:ad2} summarizes LLM-based anomaly detection studies.
Across these studies, a common pipeline is: (i) partition logs into sessions/windows, (ii) encode each sequence into a representation (templates, embeddings, or raw text), (iii) compute an anomaly signal (score/label), and (iv) optionally return \emph{evidence} (highlighted lines, retrieved neighbors, or explanations) to support triage.
Existing approaches mainly differ along four design axes: \textbf{(1) representation interface} (parsed templates vs.\ raw text), \textbf{(2) supervision setting} (fully supervised with both normal/anomalous labels vs.\ semi-supervised/one-class settings that train mostly or only on \emph{normal} data, vs.\ unsupervised settings), \textbf{(3) adaptation/training strategy} (fine-tuning/prompt-tuning vs.\ keeping the LLM frozen and using it in a training-free manner), and \textbf{(4) inference form} (discriminative classification, likelihood/perplexity scoring, prompting/ICL, retrieval, or agentic reasoning).
Notably, prompting/in-context learning (ICL) is largely orthogonal to supervision, and ``training'' is not synonymous with ``supervision'': (i) many methods \emph{train} models without using task labels (e.g., further pretraining or contrastive learning on related unlabeled logs/corpora), and (ii) some methods are \emph{training-free} but still supervised at inference time by using labeled exemplars as demonstrations in ICL.
Note that boundaries are not strict: many systems combine multiple mechanisms (e.g., encoder embeddings + prompting, or retrieval + generation).

\phead{Discriminative and representation-learning encoders.}
A large body of work uses pretrained encoders (typically BERT-like) to map log events or event sequences into embeddings, followed by a classifier or sequence model for anomaly prediction.
Representative examples that operate on parsed templates and learn discriminative decision boundaries include HitAnomaly~\cite{Huang2020HitAnomaly}, SwissLog~\cite{Li2020SwissLog}, and HilBERT~\cite{Huang2023HilBERT}.
Several studies adapt or optimize the encoder/sequence modeling pipeline, such as BERT-Log~\cite{Chen2022BERTLog}, Prog-BERT-LSTM~\cite{shao2022log}, and the semantic hierarchical compaction + BERT-based detection by Corbelle et al.~\cite{2024Semantic}.
Other works emphasize robustness and transfer across systems and unstable log vocabularies. For instance, Ott et al.\ study robustness and transferability with pretrained language models~\cite{Ott2021RobustTransferable}, and APT-LLM applies embedding-based detection to host audit and provenance event logs~\cite{benabderrahmane2025aptllm}.
Some papers explicitly investigate LLM-assisted \emph{parsing + embedding} pipelines before detection (e.g., Zhou et al.~\cite{math12172758}), while related comparisons by Thali et al.~\cite{thali2025icsllm} and Entezami et al.~\cite{entezami2025networkanomalies} evaluate ICS telemetry and network traffic without explicit runtime logs.
Finally, as LLM-era anomaly detection expands beyond classic system logs, recent large-scale studies also include LLaMA-family backbones in supervised pipelines (e.g., Horv{\'a}th et al.~\cite{horvath2025realtimelogscale}) and explanation-oriented add-ons (e.g., AnomalyExplainerBot~\cite{balasubramanian2025anomalyexplainerbot}).

\phead{Masked-LM and self-/semi-supervised normality modeling.}
A closely related family models \emph{normal} log patterns using self-supervised objectives and flags deviations as anomalies, which is well-suited to rare-anomaly regimes.
LogBert~\cite{Guo2021LogBERT} and LAnoBERT~\cite{Lee2021LAnoBERT} exemplify masked-language-model training on normal sequences, producing anomaly scores via reconstruction/prediction errors.
Recent variants continue this direction with alternative backbones and training strategies, such as LogELECTRA~\cite{Yamanaka2024LogELECTRA}, LogFiT~\cite{almodovar2024logfit}, and ADALog~\cite{Pospieszny2025ADALog}.
FastLogAD~\cite{Lin2024FastLogAD} further augments semi-supervised learning by generating pseudo anomalies to strengthen discrimination.
In practice, these methods often reduce label requirements but still rely on careful sessionization/windowing and calibration to control false positives across evolving software versions.

\phead{Reducing reliance on log parsing: from templates to raw text.}
Because parsing errors and format drift can directly harm downstream detectors, multiple works explicitly minimize or remove template parsing.
NeuralLog~\cite{Le2021NoParsing} studies anomaly detection without conventional log parsing, and LogRoBERTa~\cite{sun2026improving} further reduces parser reliance via hybrid language modeling.
In addition, several prompt- or LLM-assisted pipelines still use lightweight normalization (e.g., regex-based variable masking) as a middle ground between full template parsing and raw text, especially when deploying at scale under frequent log evolution.

\phead{Decoder-style generative forecasting and likelihood-based scoring.}
Instead of classification, decoder LMs treat anomaly detection as a \emph{language modeling} problem: a sequence is anomalous if it is hard to predict under a model trained (or adapted) on normal behavior.
LogGPT~\cite{Trabelsi2023LogGPT} is a representative GPT-style next-token prediction approach, while PreLog~\cite{jiang2024prelog} explores a pretrained model tailored for log analytics tasks (including anomaly detection) using a compact generative backbone.
Instruction-centric systems also evaluate anomaly detection within a broader log analysis interface (e.g., LogLM~\cite{liu2025loglm} and knowledge-augmented adaptation in NLPLog~\cite{adapting2025arxiv}).
For comparison, ``Good Enough to Learn'' investigates anomaly detection in ECU network packets under unreliable labels with small LMs~\cite{bogdan2025goodenough}.

\phead{Prompting and in-context learning for training-free detection and interpretation.}
With instruction-tuned LLMs, several works bypass explicit model training and instead prompt an LLM to classify sequences or explain suspicious patterns.
LogPrompt-style prompt engineering emphasizes interpretability and online usage~\cite{Liu2024LogPromptICSECompanion,Liu2024LogPromptICPC}, while feasibility studies explore ChatGPT-style models for anomaly detection under different settings and log sources~\cite{Qi2023LogGPTExploringChatGPT,Egersdoerfer2023EarlyExplorationChatGPT,mannam2023optimizing}.
Some lines treat LLMs as an \emph{operations foundation model} that supports multiple tasks, including anomaly detection, under a unified interface (e.g., OWL~\cite{owl2024iclr}).
At the benchmark level, large-context prompting is also evaluated in broad suites (e.g., LogEval~\cite{Cui2024LogEval}), clarifying how far prompting alone can go without task-specific training.

\phead{Retrieval-augmented detection: comparing against a memory of normality.}
Retrieval-based methods treat anomaly detection as \emph{nearest-neighbor comparison} in an embedding space of normal sequences and often provide evidence by returning similar historical cases.
RAPID~\cite{No2024RAPID} embeds normal windows using a pretrained encoder and flags anomalies by distance/neighborhood structure, while RAGLog~\cite{Pan2023RAGLog} frames detection through retrieval-augmented generation by retrieving similar normal cases and judging the target’s consistency.
XRAGLog~\cite{zhang2025xraglog} further targets resource efficiency and context awareness in RAG-style pipelines.
This family is especially compatible with semi-supervised settings (normal-only data) and naturally supports operator-facing justifications through retrieved exemplars.

\phead{Hybrid and collaborative frameworks: small models for scoring, LLMs for enrichment, verification, or rules.}
Many practical designs combine a lightweight detector with an LLM component that improves robustness or interpretability.
LLMeLog~\cite{he2024llmelog} enriches log events with LLM-derived semantics before training a smaller classifier. LogFormer~\cite{Guo2024LogFormer} combines representation learning with LLM assistance, and LogLLM~\cite{Guan2024LogLLM} explores joint use of encoders with larger LMs.
AdaptiveLog~\cite{ma2025adaptivelog} explicitly frames anomaly detection as collaboration between large and small models, while CoLA~\cite{zhu2025cola} studies model collaboration at scale with multiple LLMs.
Knowledge and control signals are also injected via rules or domain constraints (e.g., LogRules~\cite{huang-etal-2025-logrules}), and tuning procedures aim to stabilize performance (e.g., SemiSMAC~\cite{sun2025semismac} and SemiRALD~\cite{sun2025semirald}).
Finally, cross-system generalization and ``new system'' adaptation are increasingly important, motivating transfer-focused designs such as LogSynergy~\cite{sui2025BridgingtheGap}.

\phead{Agentic and multi-step reasoning for complex security/incident contexts.}
A growing line of work frames anomaly detection as a multi-step investigation that iteratively retrieves context, applies tools, and produces both predictions and narratives.
Examples include multi-agent collaboration for threat/anomaly scenarios (Audit-LLM~\cite{Song2024AuditLLM}), consistency-based detection with evidence building (WebNorm~\cite{liao2024detecting}), and recursive/agentic pipelines for context-aware detection and response (LogRESP-Agent~\cite{2025LogRESP-Agent}).
Related systems emphasize turning alerts into operator-facing intelligence (SHIELD~\cite{sun2025alerts2intelligence}) or improving reasoning structure and controllability (LogReasoner~\cite{ma2025logreasoner}), while insider-threat anomaly detection appears as another domain where agentic reasoning is natural (Song et al.~\cite{song-etal-2025-confront}).
Agentic methods can improve interpretability and reduce analyst workload, but introduce additional latency/cost and require careful guardrails for reliable deployment.

\phead{Benchmarks and data generation.}
As the space expands, benchmarks and data-centric efforts have become important for fair comparison and stress-testing.
Karlsen et al.\ benchmark LLMs for log analysis tasks including anomaly detection~\cite{Karlsen2024BenchmarkingLLMsJNSM}, and LogEval provides a broader benchmark suite for LLM-based log analysis~\cite{Cui2024LogEval}.
For comparison, AD-LLM evaluates LLM-based anomaly detection across general-text domains~\cite{yang-etal-2025-ad}, helping compare anomaly-detection capabilities outside runtime-log settings.
Complementarily, AnomalyGen~\cite{li2025anomalygen} explores LLM-driven generation of anomalous log sequences, reflecting an emerging direction where synthetic anomalies and augmentation are used to improve coverage of rare patterns and unstable logs.

\rqboxc{Across LLM-based anomaly detection studies, the main shift is not just higher accuracy. It is also a move toward treating anomalies as \emph{evidence-grounded inconsistencies} with expected behavior, such as deviations from retrieved ``normal'' cases, violations of explicit rules, or unusually high calibrated anomaly scores. The strongest systems therefore constrain LLMs with stable interfaces (templates, embeddings, or normalized text) and use them mainly to explain, verify, or enrich lightweight detectors. LLMs do not remove the ambiguity of what counts as an anomaly. Instead, they shift effort from feature design to normality curation, context selection (windowing or retrieval), and operational guardrails that control false alarms and cost. This makes hybrid, evidence-driven workflows a practical deployment pattern.}

\subsection{Failure Prediction}
\label{subsec:logfp}
\subsubsection{Task definition}
Log-based failure prediction is a proactive early-warning task: it forecasts whether the system (or a component) will experience a \textit{future failure} within a horizon, so operators can take preventive actions (e.g., mitigation, rollback, or resource reallocation) before the system enters an unrecoverable or user-visible failure state.
Given logs observed up to the current time $t$ (often represented as a session/window $S_{t-L:t}$ or a prefix of an ongoing session),
the task outputs a probability $\Pr(\text{failure in }[t, t+H])$ (or a binary label) for a future horizon $H$,
and sometimes additional signals such as predicted time-to-failure or early-warning decisions under different lead-time budgets.

Compared with anomaly detection, failure prediction is explicitly \textit{forward-looking} and tied to an operational failure definition
(e.g., crash, outage, severe SLO violation, incident ticket).
Its difficulty depends on the prediction horizon: useful early signals in logs may be weak, sparse, and distributed across components or long time ranges, while short-horizon signals can be clearer but less actionable.
Depending on the environment, prediction may focus on (i) failures of similar components in homogeneous systems (sequence-centric modeling),
or (ii) failures driven by interactions among heterogeneous components (leveraging cross-component correlations).

\subsubsection{LLM-based Approaches for Failure Prediction}
\label{subsec:logfp_llm}
Compared with anomaly detection, failure prediction must explicitly optimize lead time and horizon awareness:
a useful predictor should raise early warnings sufficiently ahead of a failure event while controlling false alarms.
Consequently, papers often report ranking/probabilistic metrics (e.g., AUC) and sometimes assess whether the model can provide actionable context (time-to-failure, likely causes).

\phead{Generative prediction of crash events and causes.}
CrashEventLLM casts failure prediction as a text generation problem:
given preceding logs, the model predicts forthcoming crash-related information (e.g., time and cause) using an instruction-tuned LLaMA backbone,
and evaluates generation quality with ROUGE-style metrics.~\cite{CrashEventLLM2024}
This formulation is attractive when operators want human-consumable explanations,
but it also raises a methodological question: lexical-overlap metrics (e.g., ROUGE) may not faithfully reflect correctness of timestamps/causal factors,
so future work may benefit from structured evaluation (e.g., exact match on failure type/time bins, and evidence grounding).

\phead{Discriminative early warning from language-model representations.}
FALL proposes a language-model-based detector that outputs an early-warning score (reported with AUC-style evaluation) to detect imminent failures in large-scale systems.~\cite{FALL2025}
In this design, the PLM (e.g., ELECTRA-style encoders) provides a robust representation of evolving log patterns,
and the predictor focuses on separating pre-failure prefixes from normal operation.
Conceptually, this parallels semi-supervised anomaly detection, but with labels aligned to future failure windows (pre-failure vs.\ non-pre-failure),
making horizon definition and data leakage control (e.g., avoiding post-failure artifacts) especially important.

\phead{Leveraging operational context beyond application logs.}
Combining LLMs and shell logs for backup-failure prediction highlights another practical pattern: operational failures are often best predicted using heterogeneous telemetry,
where shell commands, job traces, and system events provide context that is not present in application logs alone.~\cite{LLMShellBackup2025}
Here, an LLM can serve as a semantic integrator for noisy, free-form operational traces, potentially improving feature robustness and interpretability.

\phead{From anomaly detection to proactive fault tolerance.}
VMFT-LAD (Virtual Machine Proactive Fault Tolerance Using Log-Based Anomaly Detection) connects anomaly scoring with proactive mitigation for virtualized environments,
treating detected abnormal patterns as early signals to trigger fault-tolerance actions (e.g., proactive migration or resource adjustment).\cite{senevirathne2024virtual}
Although framed around anomaly detection, the operational goal aligns with failure prediction: catch the system early enough to intervene.
This line suggests a promising integration point: LLM-based detectors could be paired with decision policies (rule-based or learned) that explicitly optimize lead time, cost, and risk.

\rqboxc{
  The current failure prediction literature using LLMs is still sparse and heterogeneous.
  Most works either (i) reuse LLM representations for discriminative early warning, or (ii) exploit instruction-following to generate human-readable failure explanations.
  A key open challenge is establishing consistent evaluation that jointly measures predictive quality, timeliness, and actionability,
under realistic constraints such as log evolution, partial observability, and cross-component dependency.}

\subsection{Root Cause Analysis / Failure Diagnosis}
\label{subsec:logrca}
\subsubsection{Task definition}
Root cause analysis (RCA) / Failure diagnosis aims to identify the most likely underlying cause(s) of an observed incident and provide actionable, operator-facing evidence.
The goal is not only to name a suspect (e.g., component/change/event type), but also to \textit{justify} it with grounded signals so that time-to-mitigation can be reduced.

For RCA, results support log-specific findings in three situations: runtime logs are the main evaluated input, runtime logs are an explicit part of multimodal evidence, or the reported results explicitly involve logs in a setting where log use is optional.

Given an incident context (e.g., an anomalous time window or a set of correlated sessions),
inputs commonly include (i) logs from multiple components/services within a time range,
and optionally (ii) service topology/dependency information, deployment/change metadata, and alerts/metrics describing the incident.
The output is typically a ranked list of candidate causes (top-$k$ suspects),
together with supporting evidence such as representative templates/snippets, affected components, and (when available) an explanatory chain linking symptoms to causes.
Logs are especially useful in RCA because they preserve chronological symptom evidence, error/exception semantics, component-local context, and operator-facing evidence chains that connect observations to candidate causes.
The ``Dataset(s) type'' column in Table~\ref{tab:rca} reports whether each RCA study uses logs alone or combines logs with other operational artifacts, making the scope of multimodal inclusion explicit.

\begin{table*}[p]
  \centering
  \caption{LLM-based root cause analysis and failure diagnosis studies.
  }
  \label{tab:rca}
  \renewcommand{\arraystretch}{0.85}
  \resizebox{\textwidth}{!}{
    \begin{tabular}{p{3.3cm} c p{3cm} p{3.8cm} p{3.8cm} p{3.8cm}}
      \toprule
      \textbf{Paper} & \textbf{Train?} & \textbf{Dataset(s) type} & \textbf{Output} & \textbf{Metric(s)} & \textbf{Base Model(s)} \\
      \midrule

      \multicolumn{6}{l}{\textit{\textbf{Prompting \& In-Context Learning (ICL) for RCA / Diagnosis}}} \\
      \midrule
      Chen et al.~\cite{chen2024automatic} & Yes & Log, traces, metrics & Explanation; category & Micro and macro F1 & GPT-4 \\
      Shan et al.~\cite{shan2024face} & No & Log & Root-cause configuration location & Accuracy; FP & GPT-4 \\
      Zhang et al.~\cite{zhang2024automated} & No & Incident title and summary & Root cause & ROUGE-L, ROUGE-1, METEOR, GLEU, BERTScore, Nubia & GPT-3.5-turbo, GPT-4 \\
      InsightAI~\cite{11030043} & No & Log & Flame graph; user Q\&A chatbot & F1 & GPT-4o \\
      ScalaLog~\cite{10888670} & No & Log & Root-cause description & F1 & GPT-3.5-turbo \\
      \midrule

      \multicolumn{6}{l}{\textit{\textbf{Agentic \& Tool-Augmented RCA (Multi-step Reasoning)}}} \\
      \midrule
      RCAgent~\cite{wang2024rcagent} & No & Log, codebase & Root causes, solutions, evidence, responsibilities & METEOR, NUBIA, BLEURT, BARTScore, EmbScore & Vicuna-13B-V1.5-16K, GTE-LARGE (embedding) \\
      OPENRCA~\cite{xu2025openrca} & No & Log, traces, metrics & Root cause reasons and components & Accuracy & Claude 3.5 \\
      TAMO~\cite{zhang2025tamo} & Yes & Log, traces, metrics & Root cause localization; fault type classification & Acc@k; MiPr, MaPr, MiRe, MaRe, MiF1, MaF1 & GPT-4 \\
      Flow-of-Action~\cite{pei2025flow} & No & Log, traces, metrics & Root cause localization; fault type classification & Root cause location accuracy (LA); root cause type accuracy (TA) & GPT-4-Turbo \\
      RCLAgent~\cite{zhang2025adaptive} & No & Metrics and distributed traces & Root cause localization & Recall@k; MRR & Claude-3.5-sonnet \\
      MicroRCA-Agent~\cite{tang2025microrca} & Yes & Log, traces, metrics & Root-cause & Top@K & deepseek-v3 \\
      \midrule

      \multicolumn{6}{l}{\textit{\textbf{Knowledge / Graph / Code-Augmented Diagnosis}}} \\
      \midrule
      ForenSiX~\cite{11036790} & No & Log & Context extraction, data dependency graph, diagnosis & Accuracy & GPT-4o \\
      COCA~\cite{li2025coca} & No & Log, codebase, trace & Root-cause & BLEU-4, ROUGE-1, METEOR, Semantics, Usefulness, Exact Match, Top-K & GPT-4o \\
      AetherLog~\cite{cuiaetherlog} & Yes & Log & Root-cause category prediction & F1 & GPT-4o \\
      \midrule

      \multicolumn{6}{l}{\textit{\textbf{General Log Diagnosis: Fault Type / Resolution / Understanding}}} \\
      \midrule
      LogSage~\cite{xu2025two} & No & Log & Root causes; solutions & Precision, Recall, F1 & GPT-4o, Deepseek V3, Claude-3.7-Sonnet \\
      Herrmann et al.~\cite{herrmann2025diagnosing} & Yes & Issue description, log files, engineer communications & Root cause description & MSS & MIXTRAL-8X7B \\
      Huang et al.~\cite{huang2024demystifying} & Yes & Log & Fault-indicating description; fault-indicating parameter & Precision, Recall, F1 & UniXcoder \\
      LogLM~\cite{liu2025loglm} & Yes & Log & Log-problem-resolution & BLEU, ROUGE-1, ROUGE-2, ROUGE-L & LLaMA2-7B \\
      LogReasoner~\cite{ma2025logreasoner} & Yes & Log & Failure type & Accuracy; Weighted-F1 & Qwen2.5-1.5B \\
      Ji et al.~\cite{adapting2025arxiv} & Yes & Log & 0/1 failure detection; fault type & F1 & LLaMA-2-7B \\
      LogExpert~\cite{10.1145/3639476.3639773} & Yes & Log, StackOverflow issues & Executable resolution steps, command filling & BLEU-4, ROUGE-L & GPT-3.5-turbo, GPT-4 \\
      KnowLog~\cite{10.1145/3597503.3623304} & Yes & Log & Cause ranking & Accuracy; Weighted F1 & BERT \\
      AdaptiveLog~\cite{ma2025adaptivelog} & Yes & Log & Failure type & Recall@k & bert, chatgpt \\
      Taheri et al.~\cite{10891042} & Yes & Log & Fault classification & F1 & RoBERTa, BigBird, Flan-T5 \\
      \midrule

      \multicolumn{6}{l}{\textit{\textbf{Benchmarks \& Evaluation Suites}}} \\
      \midrule
      LogEval~\cite{Cui2024LogEval} & Yes & Log & Fault types & Accuracy; F1 & GPT-4 \\
      \bottomrule
    \end{tabular}
  }
\end{table*}

\subsubsection{LLM-based Approaches for Root Cause Analysis}
\label{subsec:logrca_llm}

Table~\ref{tab:rca} summarizes LLM-based root cause analysis (RCA) and closely related failure diagnosis studies.
These studies share a common goal: given runtime logs, sometimes combined with metrics/traces, code, tickets, or knowledge bases, produce an actionable RCA output such as \emph{(i)} a root-cause description or reason, \emph{(ii)} localization to components/configurations, and sometimes \emph{(iii)} remediation guidance and supporting evidence.
Across the studies within the log-specific scope, the main design differences are \textbf{(1) how the LLM is used} (prompting/ICL vs.\ agentic tool use vs.\ trained task model), \textbf{(2) what external context is integrated} (repositories, databases, SOPs, graphs, knowledge bases), and \textbf{(3) what the output is optimized for} (free-form explanation vs.\ structured labels/locations vs.\ resolutions).

\phead{Prompting and in-context learning for incident RCA.}
A pragmatic line treats RCA as a text understanding and summarization problem and relies on prompting/ICL to map incident artifacts to root-cause statements.
Chen et al.~\cite{chen2024automatic} perform automatic RCA for cloud incidents by consuming multiple incident signals (e.g., error logs, stack traces, socket metrics) and outputting both explanations and categories, evaluated with F1.
Zhang et al.~\cite{zhang2024automated} study ICL-based root causing for cloud incidents using incident titles and summaries and report a suite of generation-oriented metrics.
We discuss this study as a related comparison because its evaluated inputs do not expose runtime logs.
In privacy- or token-limited settings, InsightAI~\cite{11030043} frames RCA as interactive diagnosis over large private logs, producing operator-facing artifacts (e.g., flame graphs and Q\&A-style assistance) and evaluating quality with F1.
These ICL-style systems are attractive due to low engineering overhead (no task-specific training required), but their performance is often bounded by context selection, prompt robustness, and the fidelity of the provided evidence.

\phead{Agentic and tool-augmented RCA for complex systems.}
When RCA requires multi-step investigation (e.g., correlating signals, fetching relevant traces, querying databases, or checking code ownership), many works move from single-pass prompting to \emph{agentic} pipelines.
RCAgent~\cite{wang2024rcagent} is representative: it uses autonomous agents and tool augmentation over logs and external artifacts (e.g., databases and code repositories) to produce root causes, solutions, evidence, and responsibilities, and evaluates with multiple semantic similarity metrics.
OPENRCA~\cite{xu2025openrca} explicitly benchmarks the ability of agentic LLM systems to locate root causes from observability data (logs/metrics/traces), emphasizing accuracy on components and reasons.
Other methods specialize in structured outputs and fine-grained localization: TAMO~\cite{zhang2025tamo} targets cloud-native RCA with tool-assisted agents over multi-modality observations and evaluates localization via Acc@k alongside fault-type classification metrics, while Flow-of-Action~\cite{pei2025flow} incorporates SOP guidance to improve location/type accuracy.
RCLAgent~\cite{zhang2025adaptive} explores multi-agent recursion-of-thought over metrics and distributed traces, reporting ranking-style metrics (Recall@k, MRR).
It is discussed as a related comparison because runtime logs are not an explicit evaluated input.
MicroRCA-Agent~\cite{tang2025microrca} further integrates an anomaly-identification stage (training an Isolation Forest) before agentic RCA, reflecting a common practical pattern: \emph{small models for fast screening, agents for evidence-driven explanation and attribution}.
Overall, agentic approaches tend to improve completeness and traceability of RCA, but introduce added latency/cost and require careful guardrails to avoid brittle tool use or overconfident narratives.

\phead{Knowledge-, graph-, and code-augmented RCA.}
Another prominent direction strengthens RCA by explicitly modeling dependencies and injecting structured knowledge.
ForenSiX~\cite{11036790} combines context extraction with a data dependency graph for network diagnostics and reports accuracy, illustrating how graph structure can constrain reasoning and improve interpretability.
COCA~\cite{li2025coca} augments RCA with code knowledge (logs + codebase + stack traces) to produce root-cause outputs and evaluates with both lexical and human-centric measures, reflecting the intuition that many failures are only disambiguated with code-level context.
AetherLog~\cite{cuiaetherlog} integrates knowledge graphs with LLMs for log-based RCA category prediction, evaluated with F1.
These systems highlight that, beyond raw language ability, \emph{external structure} (code, graphs, KBs) is a key ingredient for reliable RCA—especially when symptoms are ambiguous or multiple plausible causes exist.

\phead{RCA-adjacent diagnosis: fault types, resolutions, and log understanding.}
Several studies expand beyond pinpointing a cause and instead aim to complete the diagnosis loop: identifying fault types, extracting fault-indicating evidence, or generating resolutions.
Huang et al.~\cite{huang2024demystifying} extract fault-indicating descriptions and parameters from logs (precision/recall/F1), which can be viewed as a complementary capability that improves downstream RCA quality by distilling salient evidence.
LogExpert~\cite{10.1145/3639476.3639773} focuses on generating executable resolution steps grounded in external knowledge (StackOverflow), while LogLM~\cite{liu2025loglm} frames log analysis as instruction-following and outputs problem resolutions, evaluated with text generation metrics.
For broader diagnostic labeling, LogReasoner~\cite{ma2025logreasoner} predicts failure types, and Ji et al.~\cite{adapting2025arxiv} adapt LLMs to log analysis with interpretable domain knowledge for failure detection and fault typing (F1).
At the representation level, KnowLog~\cite{10.1145/3597503.3623304} enhances log understanding and supports cause ranking, and AdaptiveLog~\cite{ma2025adaptivelog} emphasizes collaboration between small and large models for failure-type identification (Recall@k).
In telecom diagnostics, Taheri et al.~\cite{10891042} study domain-tailored models for log mask prediction used for fault classification (F1), reinforcing the importance of domain adaptation in operational contexts.
Finally, in less standardized environments, Herrmann et al.~\cite{herrmann2025diagnosing} show a case study for robotics support tickets (including logs and human communications), indicating RCA often spans heterogeneous, partially structured evidence beyond pure log streams.

\phead{Benchmarks and evaluation: what ``good RCA'' means.}
As methods diversify, benchmark efforts help clarify evaluation targets.
LogEval~\cite{Cui2024LogEval} provides a broader suite for LLM-based log analysis (including fault types), using accuracy/F1, and complements RCA-focused studies by stressing comparability and coverage.
Across the table, evaluation remains fragmented. Incident RCA often uses semantic similarity metrics. Localization prefers rank/Acc@k. Categorization uses accuracy/F1, and resolution generation uses ROUGE/BLEU.
This metric diversity reflects different operational definitions of RCA (explanation vs.\ localization vs.\ remediation), and suggests that future work may benefit from task-specific, operator-aligned evaluation protocols.

\rqboxc{Across RCA studies within the log-specific scope, LLMs matter less as standalone “reasoners” and more as components that unify heterogeneous evidence (logs, metrics, traces, code, and KBs) into a structured hypothesis with supporting traces. The most reliable designs constrain generation through retrieval, tools, SOPs, or graphs. This shifts the core challenge from modeling to \emph{evidence selection, grounding, and verifiable outputs}, and makes hybrid, tool-augmented pipelines a practical pattern.}

\subsection{Log Summarization}
\label{subsec:logsum}
\subsubsection{Task definition}
\textbf{Log summarization} compresses high-volume logs into a concise, human-consumable artifact that helps engineers quickly understand \textit{what happened} and \textit{what matters}.
It is commonly used for on-call handoff, incident reports, and rapid situational awareness, where reading raw logs is too costly.

Given a set/sequence of logs associated with a session/window (often an incident window),
the task outputs a natural-language summary and/or a compact structured summary
(e.g., a short timeline of key events, prominent error templates, involved components, and extracted entities such as IDs/paths).
Two common instantiations are:
\textit{extractive} summarization (select representative lines/templates) and
\textit{abstractive} summarization (generate a narrative description),
often combined with grouping-by-time/component to keep summaries faithful and easy to scan.

\subsubsection{LLM-based Approaches for Log Summarization}
\label{subsec:logsum_llm}

Compared with anomaly detection and RCA, log summarization is one of the most \emph{LLM-native} log analysis tasks. Its primary deliverable is a human-readable artifact, and usefulness depends on semantic coherence, correct attribution, and actionable compression instead of only classification accuracy.
Log summarization differs from generic text summarization because logs are noisy, temporally dense, uneven in informativeness, and useful only when the summary remains faithful to concrete evidence.
Across the collected studies, LLM-based log summarization typically follows a ``\emph{structure-then-generate}'' pattern: (i) pre-organize raw logs (by time, component, severity, or templates), (ii) select salient evidence (representative lines, errors, entities), and then (iii) generate an abstractive narrative or a structured synopsis (timeline, key events, involved modules), often with faithfulness constraints.

\phead{Benchmarks and evaluation: summarization as a first-class log task.}
LogEval~\cite{Cui2024LogEval} explicitly includes \emph{log summarization} in a broader benchmark suite for LLM-based log analysis.
It evaluates summary quality using both lexical overlap (e.g., ROUGE) and task-oriented correctness signals (e.g., accuracy/F1-style measures), and compares strong proprietary models against a wide range of open-source chat/instruction models.
This benchmark perspective highlights two recurring issues in summarization: (i) \emph{metric mismatch}---ROUGE-like scores only partially reflect incident usefulness, and (ii) \emph{context limitation}---performance depends heavily on which log subset is provided to the model, motivating retrieval or guided selection before generation.

\phead{Guided enhancement assistants: making summaries faithful and operator-friendly.}
Practical summarization assistants often aim to reduce hallucinations and improve scanability by enforcing intermediate structure.
logSage~\cite{logsage2025} (SAC'25) exemplifies this direction by treating summarization as an assistant workflow with guided enhancement: instead of asking an LLM to narrate raw logs end-to-end, the system emphasizes staged processing (e.g., evidence selection, grouping, and controlled rewriting) so that the final summary better reflects the underlying log evidence and aligns with on-call needs (handoff, incident notes, rapid situational awareness).
Such designs typically trade some linguistic freedom for higher faithfulness and better operator trust.

\phead{Instruction-based log interpretation as summarization.}
Some works position summarization as part of a unified \emph{instruction-following} interface for log analysis.
LogLM~\cite{liu2025loglm} transforms multiple log tasks into instruction-based interactions. For summarization-like requests, it outputs natural-language interpretations and resolutions, and evaluates with standard generation metrics (BLEU/ROUGE variants).
We discuss its summarization assignment as a related comparison because the evaluated output interprets or resolves individual log messages and does not compress a log set, sequence, session, or window.
Similarly, Ji et al.~\cite{adapting2025arxiv} emphasize interpretable domain knowledge to adapt LLMs for log analysis. Their ``interpretation'' outputs overlap with summarization in practice by explaining what the logs indicate, and are assessed via manual scoring, reflecting that human utility and correctness are hard to capture with automatic metrics alone.
This line suggests that in real deployments, summarization is rarely an isolated task. It often co-occurs with Q\&A, diagnosis hints, and ``what should I do next'' guidance.

\phead{Tooling for scalable log processing and its relationship to summarization.}
While not a summarization method per se, LogLead~\cite{mantyla2024loglead} illustrates an important enabling trend: scalable pipelines that load, enhance, and process logs efficiently can serve as the \emph{front-end} to LLM summarization by curating what the model sees (e.g., enhancing logs, selecting windows, and organizing content).
In other words, summarization quality is frequently bounded by upstream log preparation and selection instead of purely by the decoder’s generation capability.

\rqboxc{Across LLM-based log summarization studies, the key contribution of LLMs is turning noisy, heterogeneous log streams into coherent narratives and structured incident briefs. The most effective systems do not summarize ``raw logs'' directly. They first impose structure through grouping, salience selection, and evidence extraction, and then generate a constrained summary. This shifts the main challenge from generation to \emph{faithful context curation and evaluation} under operator-centric criteria.}

%% file: Cross_Cutting_Insights.tex
\section{Cross-Cutting Insights and Future Directions}
\label{sec:cross_cutting}

The preceding sections reviewed LLM-based work across the log-analysis pipeline, from logging statement generation to parsing and downstream operational tasks.
Our findings suggest that LLM4Log should not be understood as a collection of isolated task-specific upgrades.
Instead, the same challenges appear across tasks: semantic flexibility versus controllability, richer evidence integration versus higher cost and latency, and more operator-friendly outputs versus a greater need for grounding and verification.
This section summarizes these recurring patterns and highlights the most important open directions for reliable real-world adoption.

\subsection{Cross-cutting Insights and Trade-offs}
\label{subsec:cross_task_insights}

Table~\ref{tab:llm4log-paradigm-tradeoffs} summarizes the comparison by linking recurring challenges to the main method families and their practical use. In practice, this means no single paradigm dominates across the whole workflow. Different tasks and constraints favor different combinations of supervision, retrieval, control, and system design.

The first cross-cutting observation is that \emph{log-centric} LLM systems are distinct from broader LLM4AIOps settings because logs are both semantically rich and operationally noisy.
Unlike metrics or traces, logs often contain the chronological symptom evidence that operators read directly during debugging and incident triage, but they also mix free-form descriptions with identifiers, stack frames, configuration keys, and other code-like tokens.
This makes log analysis not only a pattern-recognition problem, but also an evidence-integration and decision-support problem where the quality of the final operator-facing output matters as much as the underlying prediction~\cite{Cui2024LogEval,Karlsen2024BenchmarkingLLMsJNSM,owl2024iclr}.

The second recurring insight is that LLMs are most beneficial when the main challenge is semantic variation under drift, not stable repeated syntax.
In logging-related tasks, this appears when developers need help generating or revising informative logging statements across evolving codebases and conventions~\cite{fu2014wheredolog,zhu2015learningtolog,Li2023LogMessageReadability,Rong2023LoggingPractices}.
In parsing, it appears when template boundaries are ambiguous, delimiters are unstable, or previously unseen messages emerge under version and deployment drift~\cite{llmparser2024icse,beck2025system,xu2024divlog,li2024lilac}.
In downstream tasks, it appears when semantically related failures surface through heterogeneous wording across components, or when weak signals must be integrated across long sequences instead of being recognized from a single fixed pattern~\cite{owl2024iclr,Cui2024LogEval,Karlsen2024BenchmarkingLLMsJNSM,ma2025logreasoner}.
Across these settings, the main value of LLMs is not that they eliminate preprocessing, but that they tolerate variation better than approaches tightly coupled to historically observed surface forms~\cite{llmparser2024icse,xu2024divlog,Karlsen2024BenchmarkingLLMsJNSM}.

The third pattern is that successful LLM4Log systems rarely rely on unconstrained, end-to-end generation alone.
Across tasks, the most practical designs are selective and layered: they first reduce the search space through preprocessing, retrieval, filtering, grouping, or lightweight scoring, and only then use a stronger LLM for semantic interpretation, explanation, or reporting~\cite{llmparser2024icse,owl2024iclr,ma2024luk,ma2025logreasoner}.
This pattern is visible in parsing researches that combine grouping and cache reuse with selective LLM calls, in anomaly detection pipelines that let smaller encoders or classical models narrow candidate windows before explanation, and in RCA or summarization systems that retrieve historical incidents, runbooks, or correlated artifacts before generating hypotheses or reports~\cite{llmparser2024icse,owl2024iclr,ma2024luk,ma2025logreasoner,Cui2024LogEval}.
The broader lesson is that LLMs are most useful as a high-value reasoning and explanation component inside a larger pipeline. They are not a drop-in replacement for every stage~\cite{owl2024iclr,Cui2024LogEval,Karlsen2024BenchmarkingLLMsJNSM}.

These observations also clarify the main trade-offs across current LLM4Log paradigms.
Fine-tuning is attractive when task boundaries are stable and enough supervision exists, but it is harder to refresh under rapid drift~\cite{gururangan2020dontstoppretraining,ouyang2022instructgpt}.
Prompting and ICL support quick adaptation, yet they shift the burden toward prompt engineering and context management~\cite{brown2020language,liu2021pretrainpromptpredict,min2022rethinkingicl}.
Retrieval-augmented approaches often provide a better balance for log-centric workflows because system-specific knowledge can be refreshed without retraining, but this depends on having a trustworthy retrieval layer~\cite{lewis2020rag,owl2024iclr,ma2024luk,Cui2024LogEval}.
Agentic workflows extend this flexibility further, although the added coverage comes with higher latency, system complexity, and more opportunities for cascading errors~\cite{yao2022react,mialon2023augmentedlms,ma2025logreasoner}.
Across the corpus, the most deployment-friendly compromise is often a hybrid design in which smaller models or structured filters handle routine narrowing, while a stronger LLM is invoked selectively for ambiguous cases, explanation, or cross-source analysis~\cite{Karlsen2024BenchmarkingLLMsJNSM,ma2024luk,ma2025logreasoner}.
In comparative terms, fine-tuning offers the strongest task-specific stability when labels and task definitions are reliable, but it is the least convenient to refresh under rapid drift. Prompting and ICL are easier to adapt quickly, but they are more sensitive to prompt construction and context quality. Retrieval-augmented methods improve freshness and grounding by injecting system-specific evidence at inference time, but they depend on robust indexing and retrieval quality. Agentic or hybrid designs are the most capable of integrating multi-step evidence across tools and artifacts, but they also carry the highest latency, coordination overhead, and error-propagation risk.

The ratings in Table~\ref{tab:llm4log-paradigm-tradeoffs} are design-level comparisons, not measured cross-paper scores.
We assess expected runtime cost from the number of large-model inference steps on the routine path: low indicates that a large model is invoked only for selected cases, medium indicates one main inference step per case, and high indicates repeated or multi-step model and tool calls.
Training, index-maintenance, and other update costs are excluded from the runtime rating and noted qualitatively where relevant because the reviewed studies do not report them consistently.
Drift adaptability is low when accommodating new log formats or system knowledge generally requires new labels or weight updates, medium when prompts, examples, or workflow settings can be changed without retraining, and high when system-specific evidence can be refreshed through an external index or routing layer.
Grounding support is low when outputs rely mainly on the prompt and model parameters, medium when structured context or intermediate checks are available, and high when outputs can be explicitly linked to retrieved records or logged tool evidence.
Compound ratings indicate that the level depends on the implementation.
The comparison assumes a comparable model scale and workload and does not rank every paper within a paradigm.

The final cross-task insight concerns operator alignment.
Several downstream tasks, especially RCA and log summarization, require outputs that humans can inspect, challenge, and act on.
This creates a stronger requirement than predictive accuracy alone.
A detector that produces a strong F1 score but cannot explain which evidence matters may still have limited operational value, whereas an RCA assistant or summarizer that produces plausible but weakly augmented narratives can actively mislead users.
Accordingly, many of the most promising LLM4Log directions couple generation with structured extraction, retrieved evidence, tool support, or schema-constrained outputs so that the resulting artifacts remain inspectable and verifiable instead of purely free-form~\cite{Cui2024LogEval,Karlsen2024BenchmarkingLLMsJNSM,ma2024luk,ma2025logreasoner}.

\begin{table*}[t]
\centering
\small
\renewcommand{\arraystretch}{1.15}
\setlength{\tabcolsep}{3pt}
\caption{Design-level trade-offs among major LLM adaptation paradigms in LLM4Log.}
\label{tab:llm4log-paradigm-tradeoffs}
\begin{tabular}{@{}>{\RaggedRight\arraybackslash}p{0.13\textwidth}
                >{\RaggedRight\arraybackslash}p{0.20\textwidth}
                >{\RaggedRight\arraybackslash}p{0.20\textwidth}
                >{\RaggedRight\arraybackslash}p{0.17\textwidth}
                >{\RaggedRight\arraybackslash}p{0.20\textwidth}@{}}
\toprule
\textbf{Paradigm} &
\textbf{Strength} &
\textbf{Limitation} &
\textbf{Expected runtime cost / drift adaptability / grounding support} &
\textbf{Best fit} \\
\midrule

Fine-tuning &
Stable task-specific behavior &
Needs labels and is harder to refresh &
Medium, excluding training and update cost / Low / Low &
Stable tasks with labeled data \\

\midrule

Prompting / ICL &
Fast adaptation without retraining &
Sensitive to prompts, examples, and context selection &
Medium, with context-length dependence / Medium / Low--medium &
Low-label and fast-changing settings \\

\midrule

Retrieval-augmented methods &
Fresh system-specific evidence &
Depends on retrieval quality and index freshness &
Medium, including retrieval / High with a maintained index / High when retrieved evidence is exposed &
RCA, summarization, and incident support \\

\midrule

Agentic workflows &
Multi-step evidence gathering and tool use &
More complex, with higher latency and error-propagation risk &
High / Medium--high when tools and knowledge can be updated / Medium--high when tool traces are exposed &
Complex multi-source investigation workflows \\

\midrule

Hybrid / selective escalation &
Balances efficiency and capability &
Requires routing, caching, and validation design &
Low--medium on the routine path / High with updated routing and external knowledge / Medium--high when evidence or validation traces are exposed &
Production pipelines with cost or latency constraints \\

\bottomrule
\end{tabular}
\end{table*}

\subsection{Open Challenges and Future Directions}
\label{subsec:future_agenda}

Despite rapid progress, the current LLM4Log literature still exhibits substantial limitations that make direct cross-paper comparison and real-world translation difficult.
A central issue is comparability.
Many papers report the same headline metrics, but the underlying setups often differ in ways that materially affect results: datasets differ, sessionization or windowing strategies vary, some pipelines depend on parsing while others operate on raw or lightly normalized text, anomaly ratios are inconsistent, and prompt or model settings are frequently under-reported~\cite{Cui2024LogEval,Karlsen2024BenchmarkingLLMsJNSM,beck2025system}.
These differences make it difficult to compare performance numbers across studies. 
For example, two papers may both report F1, AUC, ROUGE, or Acc@k, but those scores may be based on different label definitions, windowing or grouping strategies, anomaly ratios, parsing choices, prompt templates, model versions, decoding settings, or evaluation targets.
These differences also make performance claims harder to interpret. 
A label may define an anomaly, root cause, or correct answer differently across datasets. 
Windowing and grouping can change what evidence is available to the model and can also change the class balance. 
In highly imbalanced anomaly-detection datasets, a single aggregate score may hide important failure cases. 
Similarly, results based only on public benchmarks may overstate how ready a method is for industrial use.
For this reason, simple pooled rankings or broad claims about the universally best LLM4Log method would be misleading.
What the literature currently supports more reliably is a pattern-level summary of recurring design choices, strengths, evaluation gaps, and failure modes, not universal leaderboard conclusions~\cite{Cui2024LogEval,Karlsen2024BenchmarkingLLMsJNSM}.

The second challenge is evaluation realism.
A large fraction of the literature still depends on a relatively small set of public benchmarks, especially HDFS, BGL, and a few related datasets~\cite{he2021survey,Cui2024LogEval,Karlsen2024BenchmarkingLLMsJNSM}.
These benchmarks are useful for controlled comparison, but they only partially reflect the operational realities that motivate LLM-based log analysis in practice: long causal chains, mixed-source streams, privacy constraints, rapidly evolving software versions, and the need to correlate logs with tickets, runbooks, metrics, and code~\cite{Yang2023EmbeddedLogs,Ma2025PractitionerExpectationsLogAD,owl2024iclr}.
The corpus nevertheless provides limited and uneven evidence about operational realism.
Most studies remain benchmark-oriented.
Some studies use industrial or proprietary data, report deployment-style measures, or include human evaluation, but these forms of evidence answer different questions and should not be read as a single maturity scale.
Industrial or proprietary data can improve operational realism, but they do not establish integration into an organizational workflow.
Measures such as latency and token use provide evidence about technical feasibility, but they do not show whether operators benefit.
Human evaluation may be limited to annotation or expert rating and need not reflect sustained use.
The task distribution also suggests that readiness is task-specific.
RCA contains the clearest examples of industrial and deployment-oriented evaluation, while parsing, anomaly detection, and summarization remain largely benchmark-driven.
This distribution shows why progress in one task should not be generalized across the pipeline.
Hence, we assess practical readiness through three separate questions: whether a method works on representative operational data, whether it functions within a real workflow, and whether it helps the people who use its output.

The third challenge concerns grounding, privacy, and deployment practicality.
Operator-facing outputs in RCA and summarization can appear coherent even when they are only weakly supported by the logs or retrieved evidence, making grounding and evidence attribution essential~\cite{Cui2024LogEval,Karlsen2024BenchmarkingLLMsJNSM,ma2024luk,ma2025logreasoner}.
At the same time, logs may contain sensitive identifiers, internal topology, or security-relevant content, which complicates direct use of remote API-based LLMs and pushes many practical designs toward self-hosted, retrieval-bounded, or hybrid architectures~\cite{nist80092logmgmt,gdpr2016,nist80053r5,greshake2023indirectpromptinjection}.
These constraints are not secondary deployment details. They directly shape what kinds of LLM4Log systems are feasible in practice~\cite{gdpr2016,nist80053r5,mialon2023augmentedlms}.

Looking ahead, several research directions appear particularly important.
First, benchmark design should better reflect drift, long-tail behavior, and cross-system variation instead of relying mainly on static splits~\cite{llmparser2024icse,xu2024divlog,Cui2024LogEval}.
Second, privacy-preserving and self-hosted LLM4Log deserves more attention, including parameter-efficient adaptation, on-prem retrieval, and redaction-aware pipelines~\cite{dettmers2023qlora,gdpr2016,nist80053r5}.
Third, RCA and summarization should move toward more augmented and verifiable outputs, with explicit evidence attribution and stronger evaluation of faithfulness instead of plausibility alone~\cite{Cui2024LogEval,Karlsen2024BenchmarkingLLMsJNSM,ma2024luk,ma2025logreasoner}.
Fourth, long-context compression, memory, and adaptive evidence selection remain central challenges because incident-relevant signals are often sparse and distributed across long noisy histories~\cite{owl2024iclr,mialon2023augmentedlms,pope2023vllm}.
Fifth, the field needs broader industrial-scale evaluation beyond the standard public benchmarks, including richer multimodal settings and operator-centered studies~\cite{Yang2023EmbeddedLogs,Ma2025PractitionerExpectationsLogAD,Cui2024LogEval}.
Finally, human-in-the-loop support is likely to be more practical than full autonomy in many settings: future systems should help engineers inspect evidence faster, compare likely causes, summarize timelines, and decide where to look next while remaining transparent about uncertainty and evidence provenance~\cite{Yang2023EmbeddedLogs,Ma2025PractitionerExpectationsLogAD,ma2024luk}.

%% file: conclusion.tex
\section{Conclusion}
\label{sec:conclusion}

This survey presented a systematic review of \emph{LLM-based log analysis} across the log-analysis pipeline, covering logging statement generation and maintenance, log parsing, and downstream tasks such as anomaly detection, failure prediction, root cause analysis, and log summarization. Across these areas, the literature shows that LLMs offer clear advantages in semantic generalization, evidence integration, and operator-facing assistance, while also introducing important challenges in evaluation, grounding, privacy, cost, and deployment practicality. As discussed in Section~\ref{sec:cross_cutting}, progress in LLM4Log will depend not only on stronger models, but also on more rigorous cross-paper evaluation, better grounding and verification for operator-facing outputs, and system designs that remain reliable under real-world operational constraints.